\documentclass[twocolumn]{aastex63}

\usepackage{graphicx}
\usepackage[T1]{fontenc}
\usepackage{aecompl}
\usepackage[]{amsmath}
\usepackage{color}
\usepackage{xspace}
\usepackage{soul}

\newcommand{\msun}{\mathrm{M}_\odot}
\graphicspath{{figures/}}

\newcommand{\ud}{\mathrm{d}}

\def\lsim{ \lower .75ex \hbox{$\sim$} \llap{\raise .27ex \hbox{$<$}} }

\def\vxm{\overline{v_{\mathrm{x',i}}}}
\def\vym{\overline{v_{\mathrm{y',i}}}}
\def\vzm{\overline{v_{\mathrm{z',i}}}}

\def\vsqxm{\overline{v^2_{\mathrm{x',i}}}}
\def\vsqym{\overline{v^2_{\mathrm{y',i}}}}
\def\vsqzm{\overline{v^2_{\mathrm{z',i}}}}

\def\vsqxym{\overline{v^2_{\mathrm{x'y',i}}}}
\def\vsqxzm{\overline{v^2_{\mathrm{x'z',i}}}}
\def\vsqyzm{\overline{v^2_{\mathrm{y'z',i}}}}

\submitjournal{ApJ}

\shorttitle{Testing \textsc{jam} with \textsc{Auriga}}
\shortauthors{Wang et al.}
\graphicspath{{./}{figures/}}

\begin{document}


\title{Is the core-cusp problem a matter of perspective: Jeans Anisotropic Modeling against numerical simulations}

\correspondingauthor{Wenting Wang}
\email{wenting.wang@sjtu.edu.cn}


\author[0000-0002-5762-7571]{Wenting Wang}
\affiliation{Department of Astronomy, Shanghai Jiao Tong University, Shanghai 200240, China}
\affiliation{Shanghai Key Laboratory for Particle Physics and Cosmology, Shanghai 200240, China}
\author{Ling Zhu}
\affiliation{Shanghai Astronomical Observatory, Chinese Academy of Sciences, 80 Nandan Road, Shanghai 200030, China}
\author[0000-0001-7890-4964]{Zhaozhou Li}
\affiliation{Centre for Astrophysics and Planetary Science, Racah Institute of Physics, The Hebrew University, Jerusalem, 91904, Israel}
\author{Yang Chen}
\affiliation{Anhui University, Hefei 230601, China}
\affiliation{National Astronomical Observatory, Chinese Academy of Sciences, Datun Road 20A, Beijing 100101, China}
\author[0000-0002-8010-6715]{Jiaxin Han}
\affiliation{Department of Astronomy, Shanghai Jiao Tong University, Shanghai 200240, China}
\affiliation{Shanghai Key Laboratory for Particle Physics and Cosmology, Shanghai 200240, China}
\author{Feihong He}
\affiliation{Department of Astronomy, Shanghai Jiao Tong University, Shanghai 200240, China}
\affiliation{Shanghai Key Laboratory for Particle Physics and Cosmology, Shanghai 200240, China}
\author{Xiaohu Yang}
\affiliation{Department of Astronomy, Shanghai Jiao Tong University, Shanghai 200240, China}
\affiliation{Shanghai Key Laboratory for Particle Physics and Cosmology, Shanghai 200240, China}
\author[0000-0002-4534-3125]{Yipeng Jing}
\affiliation{Department of Astronomy, Shanghai Jiao Tong University, Shanghai 200240, China}
\affiliation{Shanghai Key Laboratory for Particle Physics and Cosmology, Shanghai 200240, China}
\author{Carlos Frenk}
\affiliation{Institute for Computational Cosmology, Department of Physics, Durham University, South Road, Durham DH1 3LE, UK}
\author{Jialu Nie}
\affiliation{National Astronomical Observatory, Chinese Academy of Sciences, Datun Road 20A, Beijing 100101, China}
\author{Hao Tian}
\affiliation{National Astronomical Observatory, Chinese Academy of Sciences, Datun Road 20A, Beijing 100101, China}
\author{Chao Liu}
\affiliation{National Astronomical Observatory, Chinese Academy of Sciences, Datun Road 20A, Beijing 100101, China}
\author{Yanan Cao}
\affiliation{National Astronomical Observatory, Chinese Academy of Sciences, Datun Road 20A, Beijing 100101, China}
\author{Xiaoqing Qiu}
\affiliation{National Astronomical Observatory, Chinese Academy of Sciences, Datun Road 20A, Beijing 100101, China}
\author{John Helly}
\affiliation{Institute for Computational Cosmology, Department of Physics, Durham University, South Road, Durham DH1 3LE, UK}
\author{Robert J. J. Grand}
\affiliation{Instituto de Astrof\'isica de Canarias, Calle Vía L\'actea s/n, E-38205 La Laguna, Tenerife, Spain}
\affiliation{Departamento de Astrof\'isica, Universidad de La Laguna, Av. del Astrof\'isico Francisco S\'anchez s/n, E-38206, La Laguna, Tenerife, Spain}
\author{Facundo A. Gómez}
\affiliation{Instituto de Investigaci\'on Multidisciplinar en Ciencia y Tecnolog\'ia, Universidad de La Serena, Ra\'ul Bitr\'an 1305, La Serena, Chile}
\affiliation{Departamento de Astronom\'ia, Universidad de La Serena, Av. Juan Cisternas 1200 Norte, La Serena, Chile}




\begin{abstract}
Mock member stars for 28 dwarf galaxies are constructed from the cosmological \textsc{auriga} simulation, 
which reflect the dynamical status of realistic stellar tracers. The axis-symmetric Jeans Anisotropic
Multi-Gaussian Expansion (\textsc{jam}) modeling is applied to 6,000 star particles for each system, to 
recover the underlying matter distribution. The stellar or dark matter component individually is poorly 
recovered, but the total profile is constrained more reasonably. The mass within the half-mass radius of 
tracers is recovered the tightest, and the mass between 200 and 300~pc, $M(200-300\mathrm{pc})$, is 
constrained ensemble unbiasedly, with a scatter of 0.167~dex. If using 2,000 particles and only line-of-sight
velocities with typical errors, the scatter in $M(200-300\mathrm{pc})$ is increased by $\sim$50\%. 
Quiescent Sagittarius dSph-like systems and star-forming systems with strong outflows show distinct 
features, with $M(200-300\mathrm{pc})$ mostly under-estimated for the former, and likely over-estimated 
for the latter. The biases correlate with the dynamical status, which is a result of contraction motions 
due to tidal effects in quiescent systems or galactic winds in star-forming systems, driving them out of
equilibrium. After including {\it Gaia} DR3 proper motion errors, we find proper motions can be as useful 
as line-of-sight velocities for nearby systems at $<\sim$60~kpc. By extrapolating the actual density 
profiles and the dynamical constraints down to scales below the resolution, we find the mass within 
150~pc can be constrained ensemble unbiasedly, with a scatter of $\sim$0.255~dex. In the end, we show 
that the contraction of member stars in nearby systems is detectable based on {\it Gaia} DR3 proper motion 
errors. 



\end{abstract}



\section{Introduction}
\label{sec:intro}

The growth of large and intermediate scales of cosmic structures in our Universe, such as cosmic filaments 
and the distribution of clumpy dark matter halos, can be modeled by the linear perturbation theory under 
the standard $\Lambda$ Cold Dark Matter ($\Lambda$CDM) cosmological model, which turns out to be remarkably 
successful\citep[e.g.][]{2004MNRAS.350.1153Y,2005MNRAS.362..505C,2012MNRAS.421.2904H,2015MNRAS.446.1356H,
2016MNRAS.456.2301W,2018MNRAS.475..676S}. On small scales, galaxies form through the gas cooling and 
condensation within dark matter halos \citep[e.g.][]{1978MNRAS.183..341W}. Smaller halos and galaxies 
can merge with larger halos, becoming the so-called substructures and satellite galaxies. On such small 
scales within dark matter halos, a series of challenges have been raised to the standard theory
\citep[e.g.][]{1999ApJ...522...82K,2011MNRAS.415L..40B,2020MNRAS.495.4570M}, among which one famous 
and hot debated issue is the "core-cusp" problem. Dark matter only simulations predict inner density 
slopes close to $-1$ (cusp), whereas the modeling of gas rotation curves or stellar kinematics in the 
central regions of low surface brightness galaxies, gas rich dwarfs and dwarf spheroids favor inner 
slopes close to $0$ (core), which brings in tension with the theory 
\citep[e.g.][]{1994ApJ...427L...1F,1994Natur.370..629M,2001AJ....122.2396D,2004MNRAS.351..903G,
2010AdAst2010E...5D}. The readers can also see \cite{2017ARA&A..55..343B} for a review. 

The core-cusp problem has invoked significant interests. It motivates alternative dark matter models such as 
the self-interacting dark matter\citep[SIDM;][]{2013MNRAS.430...81R,2013MNRAS.430..105P,2015PhRvD..91b3512F,
2015MNRAS.452.3650O}. Other promising solutions to the problem within the $\Lambda$CDM framework include, 
for example, the stellar feedback, that dark matter gets heated up by stellar winds or supernovae explosions, 
which drives repeated gravitational potential fluctuations and the dark matter particle orbits slowly 
expand \citep[e.g.][]{2005MNRAS.356..107R,2019MNRAS.484.1401R,2020MNRAS.491.4523F,2022arXiv220607069L,2022arXiv220101056B}. 
Modern numerical simulations have shown that this is possible
\citep[e.g.][]{2008Sci...319..174M,2012MNRAS.421.3464P,2014Natur.506..171P}. However, stellar feedback 
is only shown to be efficient for dwarfs with stellar to halo mass ratios in between $10^{-3}$ and $10^{-2}$,
whereas ultra faint dwarfs with this ratio smaller than $10^{-3.5}$ have formed very few stars, and thus 
stellar feedback seems very unlikely to be the main mechanism forming cores 
\citep[e.g.][]{2016MNRAS.456.3542T,Lazar2020,2020ApJ...904...45H,2020MNRAS.499.2912F}. In addition to 
stellar feedback, tidal effects were shown to be capable of forming cores in numerical simulations, even 
through galaxy interactions before infalling \citep[e.g.][]{2022MNRAS.510.2186G}. 

Note the core-like inner density profiles are only reported in some dwarf galaxies, and the constraints 
could still be limited by the statistical error or by model extrapolations
\citep[e.g.][]{2016MNRAS.463.1117Z,2020MNRAS.493.5825L,2021ApJ...909...20S}. 
Observation also demonstrates a strong diversity in the rotation curves, even at fixed maximum circular 
velocity \citep[e.g.][]{2019MNRAS.484.1401R,2020ApJ...904...45H,2020MNRAS.495...58S,2020JCAP...06..027K}, 
and the diversity is shown to be related to the stellar to halo mass ratio \citep{2020ApJ...904...45H}, to 
the star formation history \citep{2019MNRAS.484.1401R}, and to the surface brightness \citep[e.g.][]{2020MNRAS.495...58S}. 
However, it seems CDM hydrodynamical simulations cannot fully account for the diversity
\citep{2020MNRAS.495...58S,2020JCAP...06..027K}. 

In addition to the purpose of being served as theoretical predictions under $\Lambda$CDM framework and to be 
compared with observational constraints, modern numerical simulations are perhaps the most useful source of 
data to help validating various assumptions behind dynamical modeling methods of constraining the inner dark 
matter profiles. It is impossible to directly infer the level of systematic uncertainties behind dynamical 
models based on purely observational data, but numerical simulations provide us the ground truth to be compared 
with. For example, \cite{2011ApJ...742...20W} used two stellar populations with different half-mass radii to 
infer the dark matter densities of Sculptor and Fornax. Both galaxies are claimed to have cored dark matter 
halos. However, based on the APOSTLE (A Project Of Simulating The Local Environment) suite of hydrodynamic 
simulations of Local Group analogues, \cite{2018MNRAS.474.1398G} concluded that violations of the spherical 
symmetry can mistakenly result in best-fitting core profiles, when the truth is cuspy. In a later study of
\cite{2020MNRAS.498..144G}, a spherical Jeans modeling method is further tested using both CDM 
and SIDM models. For CDM model, they reported 50\% and 20\% of scatter for the enclosed mass in inner 
regions or in the half-mass radius of tracers, respectively. For SIDM dwarfs, their recovered mass profiles 
are biased towards cuspy dark matter distributions. \cite{2017MNRAS.469.2335C} also estimated the systematics 
of half-mass estimators developed in previous studies \citep{Wolf2010,2011ApJ...742...20W}, and reported 
intrinsic scatters of 23-25\%.

In this study, we first construct mock observations for the member stars of 28 dwarf systems, selected from 
the cosmological \textsc{auriga} suite of simulations. We then validate the discrete axis-symmetric Jeans 
Anisotropic Multi-Gaussian Expansion (\textsc{jam}) method based on the mock observational data. \textsc{jam} 
is widely used in estimating the underlying mass profiles for different types of galaxies
\citep[e.g.][]{2013MNRAS.432.1709C,2017ApJ...838...77L} and for dwarf galaxies in the Milky Way
\citep[e.g.][]{2013MNRAS.436.2598W}. In a previous study, \textsc{jam} is well tested using mock IFU data 
based on the Illustris simulation \citep{2016MNRAS.455.3680L}, and for a few mock dwarf galaxies with 
discrete data and in steady status \citep{2016MNRAS.463.1117Z}. Here our mock tracer stars and dwarf galaxy 
systems can more closely represent the dynamical status of real observed dwarf systems in the Milky Way, 
hence enabling us to investigate the performance of \textsc{jam} with realistic non-steady tracers.

We investigate not only the performance of \textsc{jam} but also the diversity in the best fits. We find that
the amount of bias is correlated with the current specific star formation rates, which is due to the difference 
in the dynamical status of quiescent and star-forming dwarfs. While radial scales most relevant to the core-cusp
problem are already below the resolution of the simulations, we extrapolate to the very inner radii to draw more
general inferences. We check the performance of \textsc{jam} by considering both the error-free data 
and the data after incorporating realistic observational errors as well as contamination by unbound stars. One
purpose of this study is to investigate whether current {\it Gaia} and future China Space Station Telescope 
(CSST) proper motions can be useful for such dynamical modelings.

In the following, we first introduce the \textsc{auriga} suite of simulations, our sample of dwarfs, mock
stars and the method of incorporating observational errors and modeling the contamination by fore/background 
or unbound stars in Section~\ref{sec:data}. The \textsc{jam} modeling approach is introduced in Section~\ref{sec:methods}.
Results are presented in Section~\ref{sec:results}, which include the overall model performance and dependence 
of the systematic biases on the star formation rate and dynamical status of the systems. We also discuss the 
results after incorporating observational errors $+$ unbound star contaminations, and we discuss results with 
or without proper motions and with smaller tracer sample size. We make connections to the core-cusp problem 
and discuss the detectability of contraction motions in member stars in Section~\ref{sec:disc}. We conclude in 
the end (Section~\ref{sec:concl}).

\section{Data}
\label{sec:data}

\subsection{Sample of dwarf systems, mock stars and mock images}
\label{sec:star}

\subsubsection{The \textsc{auriga} suite of simulations}
The sample of dwarf galaxies are constructed from the \textsc{auriga} suite of simulations \citep{2017MNRAS.467..179G}. 
Details about the \textsc{auriga} simulations can be found in \cite{2017MNRAS.467..179G} and \cite{2018MNRAS.481.1726G}. 
Here we make a brief introduction. 

The \textsc{auriga} simulations are a set of cosmological zoom-in simulations. The evolution of Milky-Way-mass systems 
are simulated and traced from redshift $z=127$ to $z=0$. They are identified as isolated halos from the parent 
dark matter only simulations of the EAGLE project \citep{2015MNRAS.446..521S}. The cosmological parameters adopted 
are from the third-year Planck data \citep{2014A&A...571A..16P} with $\Omega_m=0.307$, $\Omega_\Lambda=0.693$, 
$\Omega_b=0.048$ and $H_0=67.77\mathrm{km s^{-1} Mpc^{-1}}$. 

The simulations were conducted using the magneto-hydrodynamic code \textsc{arepo} \citep{2010MNRAS.401..791S} with 
full baryonic physics, which incorporates a comprehensive galaxy formation model and have higher resolutions than the 
parent simulation. The physical mechanisms of the galaxy formation model include atomic and metal line cooling
\citep{2013MNRAS.436.3031V}, a uniform UV background \citep{2009ApJ...703.1416F}, a subgrid model of the interstellar 
medium and star formation processes \citep{2003MNRAS.339..289S}, metal enrichment from supernovae and AGB stars
\citep{2013MNRAS.436.3031V}, feedback from core collapse supernovae \citep{2010MNRAS.406..208O} and the growth and 
feedback from supermassive black holes \citep{2005MNRAS.361..776S}. A uniform magnetic field with comoving strength 
of $10^{-14}$~G is set at redshift $z=127$, which quickly becomes subdominant in collapses halos \citep{2013MNRAS.432..176P,2017MNRAS.469.3185P}.

In this study, we use six Milky-Way-like systems from the "level 3" set of simulations. The six systems are 
named Au6, Au16, Au21, Au23, Au24 and Au27. The virial masses\footnote{The virial mass, $M_{200}$, is defined 
as the mass enclosed in a radius, $R_{200}$, within which the mean matter density is 200 times the critical 
density of the universe.} of their host dark matter halos are in the range of $1-2\times10^{12}\msun$. The 
typical dark matter particle mass is about $4\times10^4\msun$, while the average baryonic particle mass is 
about $5\times10^3\msun$.

\subsubsection{Dwarf systems}

Each of the six systems has its dwarf satellite galaxies, and in our analysis, we only use those dwarf systems 
which are less massive than $10^9\msun$ in stellar mass and also have more than 6,000 star particles. Here 
the upper limit of $10^9\msun$ is adopted to avoid including dwarf galaxies which are significantly more massive 
than classical dwarf spheroid galaxies analyzed in previous studies \citep[see e.g.][]{2020ApJ...904...45H}. 
Besides, as having been discussed by \cite{2016MNRAS.463.1117Z}, a discrete dataset with 6,000 stars is required 
to distinguish between core and cusp inner profiles. Thus we use such a large sample of stars as tracers, in order 
to control the statistical errors to be small enough, while we can focus on discussing systematic errors behind the 
model, but for part of our analysis, we will try a smaller tracer sample of 2,000 stars. Note, however, each star 
particle corresponds to a single stellar population, whose particle mass ranges from on average $\sim4,600\msun$ to 
$\sim6,500\msun$ for different simulations. A lower limit of 6,000 particles roughly corresponds to lower limits 
in the total stellar mass of $\sim10^{7.44}\msun$ to $\sim10^{7.59}\msun$. We also exclude those dwarfs whose 
dynamical spin axes and geometrical minor axes have strong mis-alignments (see explanations below in this section). 
In the end, we have 28 dwarf systems in total. For each dwarf system, we randomly pick up 6,000 bound star 
particles as tracers. Note as we have explicitly checked, different random selections of the tracer sample lead 
to differences smaller than the symbol sizes in relevant figures of this study.

Out of the 28 dwarfs, we will explicitly show the best-fitting and true density profiles for six systems similar to 
the Sagittarius dwarf spheroid (dSph)\footnote{Systems with galactocentric distances $<\sim60$~kpc and with stellar
mass in the range of $7.4<\log_{10}M_\ast/\msun<8.5$ are defined as similar to the Sagittarius dSph.}
and other four representative star-forming systems with prominent galactic winds or gas outflows\footnote{We define
systems with prominent galactic winds or gas outflows, by requiring the stellar mass in wind particles should be
greater than 15\%. In fact, most of systems with prominent galactic winds in this study have this fraction around 
30-50\%.}. Table~\ref{tbl:6sag} summarizes the information of these systems. Interestingly, only one out of the six 
Sagittarius dSph-like systems is as old as $10^{11.2}$~Gyr, whereas the other three systems still have star formations 
at $\sim$6 to 9~Gyrs ago, indicating such massive satellites can persist star formations after falling to the current 
host halo. Compared with the nearby Sagittarius dSph-like systems, other star-forming systems with strong outflows are 
at relatively large distances. This is consistent with the conclusion of \cite{2018MNRAS.478..548S}, which reported
strong mass and distance-dependent quenching signals in satellites of Local-Group-like systems from 30 zoom-in simulations.

\begin{table}
\caption{Host system name, dwarf ID, stellar mass, distance to the 
observer and the age, for six Sagittarius dSph-like systems and four star-forming systems with prominent 
gas outflows from the level 3 set of \textsc{auriga} simulations. The dwarf ID is provided in the second 
column, which is simply the position index of the dwarf in the corresponding subhalo catalog. }
\begin{center}
\begin{tabular}{lcccc}\hline
\hline
\multicolumn{1}{c}{Host name} & \multicolumn{1}{c}{dwarf} & \multicolumn{1}{c}{$\log_{10}(M_\ast/\msun)$} & \multicolumn{1}{c}{$D$[kpc]} & \multicolumn{1}{c}{age[Gyr]}\\ \hline
Au16 &  9  & 7.933 & 24.53 & 11.20 \\
Au21 &  10 & 8.446 & 42.29 & 6.75  \\
Au23 &  4  & 8.297 & 45.74 & 7.94  \\
Au23 &  7  & 8.047 & 34.52 & 8.96  \\
Au 24 & 24 & 7.503 & 60.15 & 9.36  \\
Au 27 & 25 & 7.598 & 27.09 & 9.57   \\
\hline
Au21 &  7  & 7.287 & 449.04 & 11.22 \\
Au24 &  9  & 7.842 & 235.10 & 10.11  \\
Au24 &  13  & 7.524 & 261.72 & 10.25  \\
Au27 &  3  & 8.054  & 385.27 & 9.66 \\
\hline
\label{tbl:6sag}
\end{tabular}
\end{center}
\end{table}      

\subsubsection{Mock stars}

To create mock samples of ``observed'' stars in each dwarf system, we start from the original coordinates in the 
simulation boxes and subtract the stellar mass weighted mean coordinates and velocities of all bound particles 
belonging to the dwarfs, to eliminate the perspective rotation \citep{1961MNRAS.122..433F}. We place the observer 
on the disk plane, which is defined as the plane perpendicular to the minor axis of all bound star particles with
galactocentric distances smaller than 20~kpc. The observer is 8~kpc away from the galactic center, with a random 
position angle.

The coordinates and velocities are then transformed to the observing frame. The $z'$-axis of the observing frame is 
chosen as the line-of-sight direction. The $x'$-axis (major axis) is the cross product between the spin axis of the 
dwarf galaxy and the $z'$-axis, which is projected on the ``sky''. The $y'$-axis (minor axis) is the cross product
between $z'$ and $x'$ vector, taking minus sign. We introduce the minus sign here because, according to
\cite{2013MNRAS.436.2598W}, the observing frame of the \textsc{jam} model is a left-handed system.

\subsubsection{Mock dwarf images and Multi-Gaussian Expansion}
\label{sec:imagemge}

In \textsc{jam} modeling, the potential and density distribution of the luminous stellar component will be directly 
inferred from the projected optical images of the dwarfs, with the stellar-mass-to-light ratios ($M/L$) being free 
parameters. The image will be deprojected to be in 3-dimension based on the distance and inclination angle of the 
dwarf. The inclination angle is defined as the angle between the average spin axis of the dwarf and the line-of-sight 
$z'$ direction, with the value in between 0 and 180 degrees. 

Thus to apply \textsc{jam} we need to create mock images for our sample of dwarfs. We simply adopt the 
projected stellar mass density distribution to create the images, i.e., the read in each pixel is in unit 
of $\msun/\mathrm{pc}^2$ based on all bound star particles associated to the dwarf galaxy, so in our case 
the true value of $M/L$ is unity. 

Once the mock images are made, the luminous stellar mass distribution, $\Sigma(x',y')$, will be decomposed to 
a few different Gaussian components (Multi-Gaussian Expansion or MGE in short), in order to enable the analytical
deprojection for any arbitrary $\Sigma(x',y')$ and to bring analytical solutions for any arbitrary matter distribution 
(see Section~\ref{sec:methods} for more details). In fact, the process of creating MGEs requires the $x'$-axis to 
be defined as the longer axis of the galaxy image. For ideally axis-symmetric systems, the $x'$-axis as we defined 
above is expected to be the projected longer axis as well. However, real triaxial systems might deviate from this, 
because its minor and spin axes might misalign. As we have checked, most of the dwarfs in our analysis are not 
strongly triaxial, and their image $x'$-axis is indeed aligned with the long axis, with a misaligned angle of 
at most $\sim15$ degrees. In a few cases, the $x'$-axis might be almost perpendicular to the actual long axis, 
but the galaxy images are very close to be spherical, with a very tiny difference between the major and minor axes. 
In a few very extreme cases, the dwarfs are strongly elongated, whose spin axes prominently deviate from their minor 
axes, and the $x'$-axes are very different from the image longer axis. These systems usually significantly deviate 
from axis-symmetry and are out of equilibrium (not in steady states). The performance of \textsc{jam} is expected 
to be very poor once applied to such systems \citep[e.g.][]{2016MNRAS.455.3680L}, regardless of how the $x'$-axis 
is defined. Thus we have excluded them from our analysis. Observationally, such extremely triaxial systems can also 
be excluded by comparing their image minor axis and the spin axis inferred from the line-of-sight velocity maps. 

\subsection{Incorporating observational errors}
\label{sec:err_bkgd}

\begin{figure*} 
\begin{center}
\includegraphics[width=0.8\textwidth]{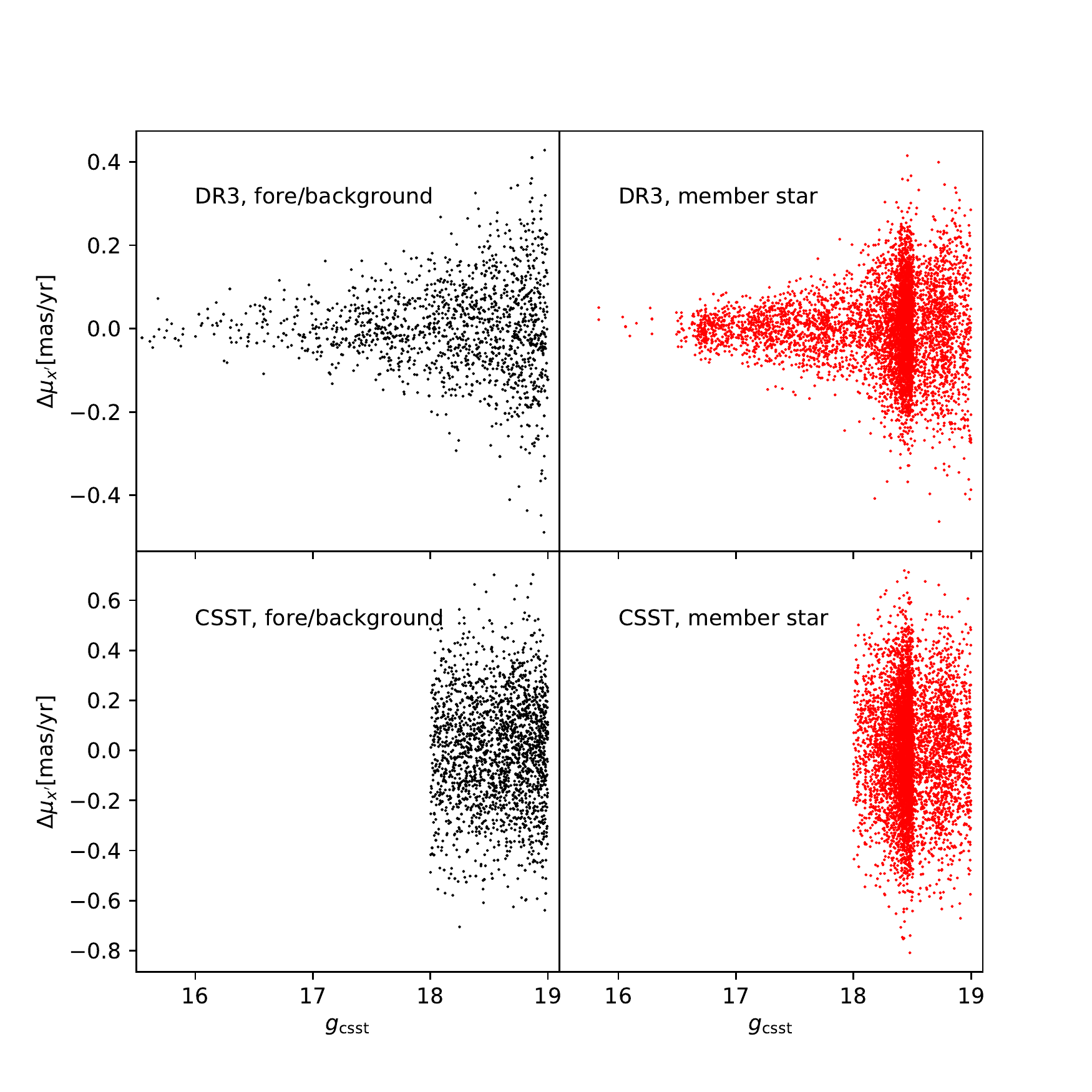}%
\end{center}
\caption{Differences between the proper motions after incorporating realistic observation errors and the true 
proper motions, reported as a function of the CSST $g$-band apparent magnitudes. We only show the $x'$ component, 
as the $y'$ component is very similar. Black (left) and red (right) points are fore/background stars and member 
stars of Au16-9 in the simulation, respectively. The proper motion errors are for either {\it Gaia} DR3 (top 
panels) or for CSST (bottom panels). The dense clumps at $g\sim18.4$ in right panels correspond 
to the red giant branch bump. {\it Gaia} DR3 proper motion errors are about a factor of two smaller than the 
upper limits in CSST errors at $18<g<19$. }
\label{fig:error}
\end{figure*}

In our analysis, we first use true positions and velocities of 6,000 bound star particles without including 
realistic errors. In this way we can better understand the intrinsic systematic uncertainties behind the model. 
We will then repeat our analysis for only 2,000 bound star particles with only line-of-sight velocities plus 
typical errors. This is for more practical meanings, because the number of observed member stars in their host 
dwarfs is at most $\sim$2,000 for current observations, most of which do not have accurate proper motion measurements.
Moreover, for the six nearby Sagittarius dSph-like systems, we will incorporate realistic observational errors 
to both line-of-sight velocities and proper motions, and model the contamination of fore/background or unbound 
stars, with member stars selected based on the differences in their kinematics relative to the dwarf centers. 
This is going to be compared with the error free and fore/background free model performance. Lastly, we find 
most Sagittarius dSph-like systems are undergoing contractions due to the effect of tidal force. We will check 
whether the amount of contraction is observable after including the {\it Gaia} DR3 proper motion errors. We 
now introduce how the observational errors are incorporated and how the fore/background is modeled.

\subsubsection{Assigning apparent magnitudes of individual stars}

Star particles in current state-of-the-art simulations represent simple stellar populations (SSP), and thus we 
cannot directly use their stellar properties, such as the magnitudes and colors, to represent single stars, 
though we can still use the position and velocity information of star particles in the simulations. To assign each 
star particle the appropriate magnitude information and hence observational errors, we use the \textsc{trilegal} 
code \citep[][Chen et al. in preparation]{2005A&A...436..895G,2009A&A...498...95V,Girardi2012,Girardi2016} to 
generate populations of different types of stars based on the input star formation history (SFH), the age-metallicity
relation (AMR) and the total stellar mass. The SFH of the Sagittarius dSph is taken from \cite{2014ApJ...789..147W}, 
and we calculate the AMR based on the closed-box model, which agrees well with real data
\citep[e.g.][]{2000AJ....119.1760L}. For the other dwarfs, the magnitude information is only required when we 
generate realistic errors for the line-of-sight velocities of mock stars (the case of using 2,000 star particles 
as tracers). For them we simply use the same SFH and AMR as the Sagittarius dSph. This is approximately a 
reasonable choice for dwarf galaxies, because dwarfs are relatively old, and the differences in their SFH and 
AMR do not lead to significant variations in the magnitude distribution of member stars.

The filter response curve is for the China Space Station Telescope \citep[CSST;][]{zhan2011,Cao2018,
Gong2019} $g$ filter. The absolute magnitudes generated by \textsc{trilegal} are converted to apparent 
magnitudes according to the distance modulus of each dwarf galaxy with respect to the mock observer in the 
simulation (see Subsection~\ref{sec:star} above). In our analysis, we will incorporate typical proper motion 
errors from either {\it Gaia} DR3 or from CSST\footnote{There are many ground based proper motion measurements 
\citep[e.g.][]{2001ApJS..133..119R,2004AJ....127.3034M,2004ApJS..152..103G,2017ApJS..232....4T,2020ApJS..248...28T,
2021MNRAS.501.5149Q}, but the typical errors are not small enough for detecting internal kinematics of dwarf
galaxies.}. When adopting the {\it Gaia} errors, we convert CSST $g$ magnitudes to {\it Gaia} $G$ magnitudes 
based on an empirical and color dependent relation linking the $g$ filter of the Hyper Suprime-Cam survey
\citep{2018PASJ...70S...1M,2018PASJ...70S...2K,2018PASJ...70S...3F} to {\it Gaia} $G$ \citep{2019MNRAS.487.1580W},
assuming CSST $g$ filter is not very different from HSC $g$ filter. This is a reasonable approximation, according 
to the CSST and HSC filter response curves \citep{Gong2019,2018PASJ...70...66K}.

The total number of stars generated by \textsc{trilegal} is very large, while the number of star particles used 
as tracers and bound to each dwarf galaxy in the simulation is more limited (in our case we use a subset of 6,000 
or 2,000 as mentioned above). We take those \textsc{trilegal} stars whose apparent magnitudes are in a certain 
range of apparent magnitudes at the corresponding distance of each dwarf, randomly select 6,000 or 2,000 
of them, and assign their magnitudes to our tracer star particles. The magnitude range is chosen as $10<g<19$ 
({\it Gaia} DR3 proper motion errors) or $18<g<19$ (CSST proper motion errors) for the Sagittarius dSph-like 
systems. These nearby systems can all have more than 6,000 stars with $g<19$. The magnitude range is $18<g<19$ 
when using CSST proper motion errors, because stars brighter than $g=18$ will be saturated in CSST observations. 
In particular, when we repeat our analysis with the smaller sample of 2,000 star particles as tracers (line-of-sight 
velocities only and with typical errors), we only use dwarf galaxies which can have more than 2,000 mock stars 
observed above the given apparent magnitude thresholds\footnote{This is based on the \textsc{trilegal} prediction, 
with the total stellar mass of the dwarf system in the simulation as the normalization. Note when we test the 
error-free case, we simply use all dwarfs, without considering whether they can have enough number of bright stars.}. 
For this case, we choose the magnitude range of $10<g<19$ for nearby Sagittarius dSph-like systems, and the range 
is chosen as $10<g<21$ for more distant systems.

\subsubsection{parallax, line-of-sight velocity and proper motion errors}

According to the apparent magnitudes, we assign each star particle errors in parallax, line-of-sight velocity and 
proper motion. The parallax errors are the median {\it Gaia} DR3 errors at the corresponding magnitude \citep{2021A&A...649A...1G,2016A&A...595A...1G}. The errors of the line-of-sight velocities are linear interpolations 
between 1 and 10~km/s, assuming stars with $g=10$ have 1~km/s of errors in the line-of-sight velocity, while stars with 
$g=21$ have the 10~km/s error. The line-of-sight velocity errors are typical for current or future spectroscopic surveys 
such as the Dark Energy Spectroscopic Instrument \citep[][DESI]{2020RNAAS...4..188A,2016ApJ...833..272Y,2022MNRAS.511.5536R}.
The errors in proper motions are either the median {\it Gaia} DR3 errors or the typical CSST errors at the corresponding
magnitudes. 

For CSST, details about how its proper motion errors are modeled and predicted will be presented in a separate 
paper (Nie et al., in preparation). Nie et al. generated mock stars in the Galactic bulge region, and realistic 
star images are created at different epochs after including PSF, noise due to correction of flat and bias, readout 
noise, sky background, parallax and stellar motions. The process of source detection and centroid position determination 
at different epochs are then modeled. After calibrating the reference frame using mock reference stars, the final 
astrometric solutions and associated errors are obtained. The reference stars are assumed to have proper motions available 
from {\it Gaia}. Details about the distribution of reference stars, relevant CSST instrument parameters, how the mock 
star samples, stellar motions and PSF are modeled can be found in Nie et al. Moreover, their PSF model is simplified, 
which might be different from the real PSF during real operations. Based on about six times of astrometric measurements 
evenly distributed in 10 years of baseline, the typical proper motion error is $\sim$0.2~mas/yr at $18<g<19$. Note 
because the number density of member stars in dwarf galaxies is less dense than the Galactic bulge region, their error 
is likely an upper limit. For {\it Gaia}, the median errors are obtained by querying the {\it Gaia} database, with typical 
DR3 proper motion error of $\sim$0.1~mas/yr at $18<g<19$.

After realistic observational errors are obtained, the proper motion, distance and line-of-sight velocity of each star 
particle are displaced from their true values according to the assigned errors, by adding a displacement drawn from a 
Gaussian distribution with the standard deviation equaling to the corresponding error. 

\subsubsection{Modelling of fore/background and unbound star contamination}
After observational errors are incorporated, we select as tracers those star particles whose differences 
in line-of-sight velocities, proper motions and parallaxes with respect to the dwarf centers\footnote{Mass 
weighted mean coordinates and velocities of all bound particles belonging to the dwarfs,} are all smaller 
than five times the corresponding errors \citep[e.g.][]{2020MNRAS.497.4162V}. In this way not only those 
true bound star particles are selected as tracers, but also some foreground and background particles as well 
as unbound particles around the dwarfs are included. To assign magnitudes for fore/background and unbound 
particles, we adopt the sample of stars created by \textsc{trilegal} for our Milky Way, and only use those 
stars in a rectangular region approximately centered on the observed Galactic latitude and longitude of the 
Sagittarius dSph ($0^\circ<l<11^\circ$ and $-20^\circ<b<-8^\circ$). We randomly select a subsample of 
\textsc{trilegal} stars and assign their magnitudes to fore/background and unbound star particles in the 
simulation. The observational errors for fore/background and unbound particles are generated in the same 
way as for bound member star particles based on their apparent magnitudes. 

Notably, for Sagittarius dSph-like systems, almost all bound star particles will be selected as tracers 
in this way, and we ensure these bound tracers to be the same as the subset used in the error-free case, 
for fair comparisons. Other selected unbound star particles are randomly picked up in proportional to 
the number of selected fraction of bound star particles. 

Figure~\ref{fig:error} shows the differences between the proper motions after assigning errors and the true 
proper motions (Au16-9), as a function of the apparent magnitudes and for both {\it Gaia} DR3 and CSST proper 
motion errors. The scatter is a factor of two smaller for {\it Gaia} DR3 errors than CSST errors at $18<g<19$, 
and is smaller than 0.1~mas/yr at $g<18$ for {\it Gaia} DR3. The dense clump of stars in the right panels and 
at $g\sim18.4$ corresponds to the red giant branch bump, whereas there are no such prominent features in the
sample of fore/background stars in the left panels. 

\section{Methodology}
\label{sec:methods}

Jeans Anisotropic Multi-Gaussian Expansion (\textsc{jam}) is a publicly available source of code\footnote{http://github.com/lauralwatkins/}. It is a powerful tool to constrain both the underlying matter distribution and the internal dynamics of 
tracers \citep[e.g.][]{2016MNRAS.462.4001Z}, based on either line-of-sight velocities or proper motions 
of tracers. The version of \textsc{jam} we use for this paper is slightly different from the public version 
of the \textsc{jam} model for discrete data \cite{2013MNRAS.436.2598W}, with improved python interface 
and plotting tools. Details about \textsc{jam} can be found in \cite{2008MNRAS.390...71C} and \cite{2013MNRAS.436.2598W}, 
and here we only briefly introduce the method. 

The method is based on solving the axis-symmetric Jeans equation in an intrinsic frame defined on the 
dwarf galaxy with cylindrical coordinates, to solve for the first and second velocity moments.

\begin{equation}
    \frac{\nu(\overline{v_R^2}-\overline{v_\phi^2})}{R}+\frac{\partial \nu \overline{v_R^2}}{\partial R}+\frac{\partial \nu \overline{v_R v_z}}{\partial z}=-\nu \frac{\partial \Phi}{\partial R}
    \label{eqn:jeans1}
\end{equation}

\begin{equation}
    \frac{\nu \overline{v_R v_z}}{R}+\frac{\partial \nu \overline{v_R v_z}}{\partial R}+\frac{\partial \nu \overline{v_z^2}}{\partial z}=-\nu \frac{\partial \Phi}{\partial z},
    \label{eqn:jeans2}
\end{equation}
where $\nu$ is the tracer density distribution. $\Phi$ is the total potential. 
Upon solving the equation to obtain unique solutions, the cross velocity terms are assumed to be zero, i.e., $\overline{v_R
v_z}=0$. In addition, the anisotropy parameter, $b$, is assumed to be constant and defined as
$\overline{v_R^2}=b\overline{v_z^2}$. A rotation parameter, $\kappa$, is introduced as $\overline{v_\phi}=\kappa
(\overline{v_\phi^2}-\overline{v_R^2})^{1/2}$. 

In our analysis, we define the $z$-axis of the intrinsic frame as the direction of the averaged spin of all bound star 
particles to the dwarf in the simulation, and the intrinsic frame is a right handed system. The intrinsic frame is linked
to the observing frame (see Section~\ref{sec:star} above) through the inclination angle, $i$, of the dwarf galaxy

\begin{equation}
    \left( \begin{array}{c}
        x' \\
        y' \\
        z'
    \end{array} \right) = \left( \begin{array}{ccc}
        1 & 0 & 0 \\
        0 & - \cos i & \sin i \\
        0 & \sin i & \cos i
    \end{array} \right) \left( \begin{array}{c}
        x \\
        y \\
        z
    \end{array} \right),
\end{equation}
and
\begin{equation}
    \left( \begin{array}{c}
        v_{x'} \\
        v_{y'} \\
        v_{z'}
    \end{array} \right) = \left( \begin{array}{ccc}
        \cos i & - \sin i & 0 \\
        \sin i & \cos i & 0 \\
        0 & 0 & 1
    \end{array} \right) \left( \begin{array}{c}
        v_R \\
        v_\phi \\
        v_z
    \end{array} \right),
\end{equation}
where $R=\sqrt{x^2+y^2}$.

The total potential, $\Phi$, on the right hand side of Equations~\ref{eqn:jeans1} and \ref{eqn:jeans2}, is contributed 
by both luminous and dark matter. As we have mentioned, the luminous matter distribution is directly inferred from the
surface brightness of the dwarf galaxy (see Section~\ref{sec:star} above). To model the density profile of dark matter,
we adopt in our analysis either the NFW profile or a double power law functional form of

\begin{equation}
    \rho(r)=\frac{\rho_s}{(r/r_s)^\gamma (1 +r/r_s)^\alpha},
    \label{eqn:double}
\end{equation}
with the model parameters ($\rho_s$, $r_s$, $\gamma$ and $\alpha$) to be constrained. Note in our analysis throughout 
this paper, the outer power law index, $\alpha$, will be fixed to 3. 

In order to have analytical solutions for any given potential model and tracer distribution, MGE is not only applied 
to the 2-dimensional surface density distribution of the luminous stellar component (see Section~\ref{sec:imagemge}
above), but also to the underlying model for the dark matter distribution and to the density distribution of tracers 
($\nu$) as well\footnote{In our case, tracers and the luminous stellar component have the same distribution, and therefore 
the same MGEs. Note the normalization of the MGE components for tracers is not important, which cancels out on two sides 
of the equations.}. Each MGE component would have analytical solutions to Equations~\ref{eqn:jeans1} and \ref{eqn:jeans2}. 
In principle, each MGE component of the tracer population can have its own rotation parameter, $\kappa_k$, and velocity
anisotropy parameter, $b_k$. $M/L$ for each MGE component can also differ, but in our analysis we treat $\kappa$, $b$ 
and $M/L$ to be the same for different MGEs. 

For an observed star with position $\boldsymbol{x'}_i=(x'_i,y'_i)$ on the image plane, which has observed velocity $\boldsymbol{v}_i=\left(v_{x',i},v_{y',i},v_{z',i}\right)$ and error matrix of 

\begin{equation}
  \boldsymbol{S}_i =
  \left( {\begin{array}{ccc}
    \sigma^2_{v_{x'},i} & 0 & 0 \\
    0 & \sigma^2_{v_{y'},i} & 0 \\
    0 & 0 & \sigma^2_{v_{z'},i} \\
  \end{array} } \right),
\end{equation}
its position, $\boldsymbol{x'}_i$, can be transformed to the intrinsic frame to solve the corresponding velocities 
and velocity dispersions, based on a set of model parameters, $\Theta$. Solution for each MGE is sought, and solutions 
of different MGEs are added together in the end. The solutions are then transformed back to the observing frame. The 
mean velocity predicted by the model in the observing frame is denoted as $\boldsymbol{\mu}_i=\left( v_{x',i},
v_{y',i},v_{z',i} \right)$, and the covariance matrix is defined through both the first and the second velocity moments

\begin{align}
    &\boldsymbol{C}_i = \nonumber\\
    &\left( \begin{array}{ccc}
        \vsqxm - \vxm^2 & \vsqxym - \vxm\,\vym & \vsqxzm - \vxm\,\vzm \\
        \vsqxym - \vxm\,\vym & \vsqym - \vym^2 & \vsqyzm - \vym\,\vzm \\
        \vsqxzm - \vxm\,\vzm & \vsqyzm - \vym\,\vzm & \vsqzm - \vzm^2
    \end{array} \right). 
\end{align}

By assuming the velocity distribution predicted by the model is a tri-variate Gaussian with mean velocity 
$\boldsymbol{\mu}_i$ and covariance $\boldsymbol{C}_i$ at $\boldsymbol{x'}_i$, the likelihood can be 
written as

\begin{align}
    L_i^\mathrm{dwarf} & = p \left( \boldsymbol{v}_i | \boldsymbol{x'}_i, \boldsymbol{S}_i, \Theta
        \right) \nonumber \\
        & = p \left( \boldsymbol{v}_i | \boldsymbol{x'}_i, \boldsymbol{S}_i, \boldsymbol{\mu}_i, \boldsymbol{C}_i
        \right) \nonumber \\
    & = \frac{ \exp \left[ - \frac{1}{2} \left( \boldsymbol{v}_i - \boldsymbol{\mu}_i
        \right)^{\mathrm{T}} \left( \boldsymbol{C}_i + \boldsymbol{S}_i \right)^{-1} \left( \boldsymbol{v}_i -
        \boldsymbol{\mu}_i \right) \right] } { \sqrt{ \left( 2 \pi \right)^{3} \left|
        \left( \boldsymbol{C}_i + \boldsymbol{S}_i \right) \right| } }.
    \label{eqn:like}
\end{align}

In addition to the above model for the dwarf galaxy itself, the discrete \textsc{jam} model can also model a population 
of fore/background stars. The surface density of fore/background stars is assumed to be constant throughout the field of 
the dwarf, and is modeled to be a given fraction, $\epsilon$, of the central surface density of the dwarf, $\epsilon \Sigma(0,0)$. 
With the assumptions, the prior of dwarf membership can be written as

\begin{equation}
    m_i(\boldsymbol{x}_i)=\frac{\Sigma(\boldsymbol{x}_i)}{\Sigma(\boldsymbol{x}_i)+\epsilon \Sigma(0,0)}.
\end{equation}

The velocity distribution of the fore/background stars can also be modeled as a tri-variate Gaussian with 
given mean velocity and velocity dispersions. Then the likelihoods of the dwarf and fore/background can 
be combined through

\begin{equation}
    L_i=m_i(\boldsymbol{x}_i)L_i^\mathrm{dwarf}+[1-m_i(\boldsymbol{x}_i)]L_i^\mathrm{bkgd}.
\end{equation}

The total likelihood is the product of the likelihood for each star

\begin{equation}
    L=\prod_{i=1}^{N_\mathrm{star}}L_i.
\end{equation}

The list of parameters used in our modeling are provided in the following:

(1) Rotation parameter, $\kappa$;

(2) Velocity anisotropy, $b$;

(3) Stellar-mass-to-light ratio, $M/L$;

(4) Dark matter halo scale density, $\rho_s$;

(5) Dark matter halo scale radius, $r_s$;

(6) Inner density slope of the host dark matter halo, $\gamma$;

(7) The background fraction, $\epsilon$.

In particular, instead of directly fitting $\rho_s$ and $r_s$, we fit $d_1=\log_{10}\rho_s^2 r_s^3$ 
and $d_2=\log_{10}\rho_s$. This is to alleviate the degeneracy between the halo parameters and also 
make the parameter space to be more efficiently explored in log space \citep{2016MNRAS.463.1117Z}. 
In our analysis, $M/L$ will mostly be fixed to unity, i.e, its true value, except for some 
cases which will be otherwise specified. The inner density slopes are also sometimes fixed to 1, i.e., 
for the NFW model profile. For fore/background contamination free cases, we fix $\epsilon$ to 0. Throughout 
this paper, we fix the distance and the inclination angle to be their true values. However, we have 
explicitly tested the results when the distance is allowed to be a free parameter. In this case, the 
best-constrained mass profiles remain very similar, while the best-fitting second velocity moments in 
very inner regions can be slightly improved, though the best-fitting distance might slightly differ 
from the true distance by up to $\sim$10\%. Moreover, the outer density slope, $\alpha$, will be fixed 
to $3$, but we have also tried to vary the outer slopes and add a constant background density, and 
our conclusions are not affected.

\section{Results}
\label{sec:results}

\subsection{Overall performance}

\begin{figure*} 
\includegraphics[width=0.49\textwidth]{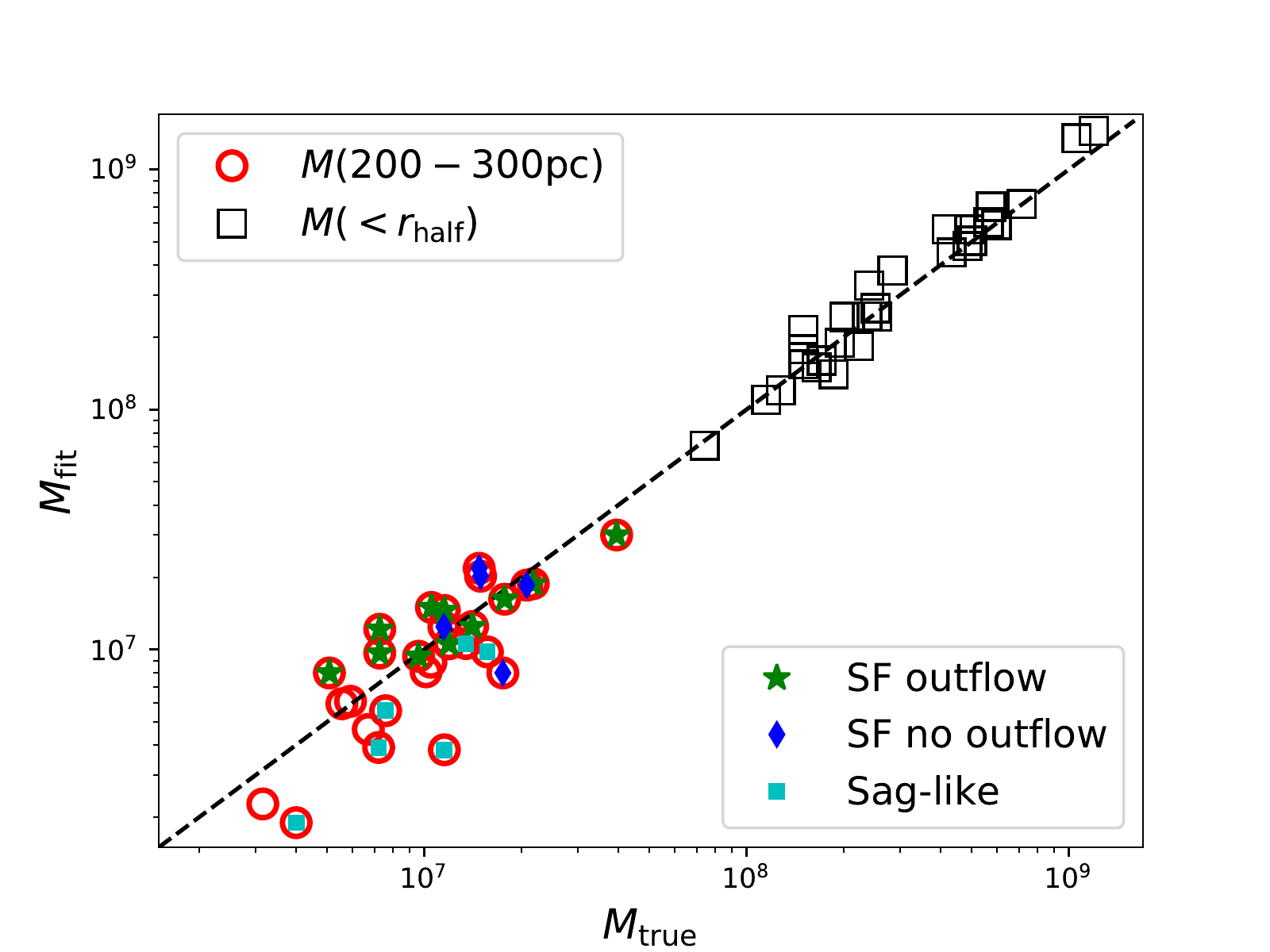}%
\includegraphics[width=0.49\textwidth]{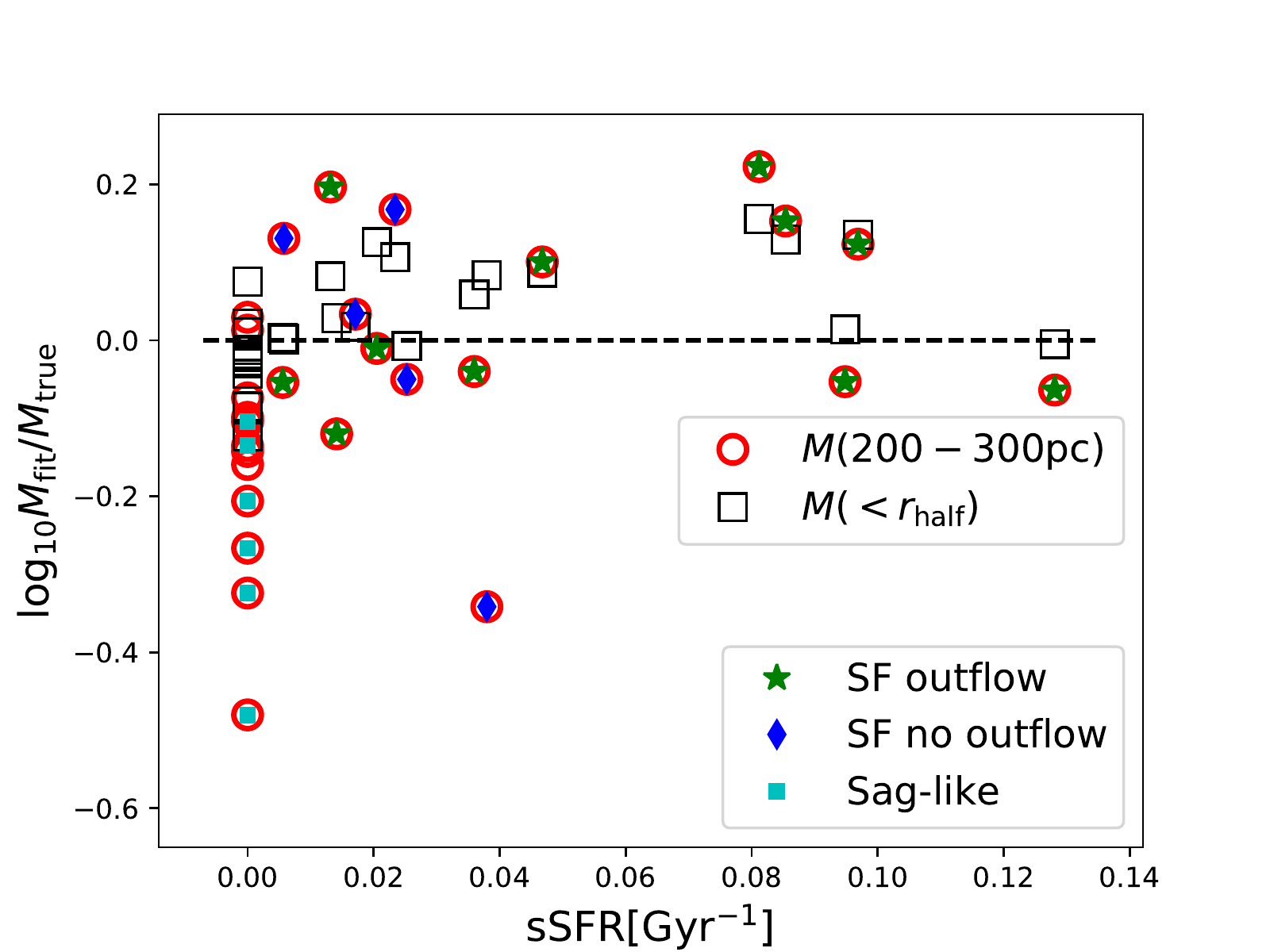}%
\caption{{\bf Left:} The best-fitting mass in between 200 and 300~pc (red circles), and within the half-mass radius
(black empty squares), versus the truth. The black dashed line marks $y=x$ to guide the eye. {\bf Right:} The
best-fitting mass in between 200 and 300~pc (red circles) and within the half-mass radius (black empty squares), 
versus the specific star formation rate (sSFR). The black dashed line marks $y=0$ to guide eye. In both plots, the 
best fits are achieved using the double power law model profile for the underlying dark matter, with free inner 
slopes, while the mass-to-light ratio, $M/L$, is fixed to the true value of unity. Only bound star particles are 
used as tracers, and no observational errors are included for results in this plot. Red circles with a green star 
symbol are star-forming systems with prominent galactic winds. Red circles with a blue diamond are systems with 
larger than zero SFR but without obvious galactic winds. Red circles with a cyan square are nearby Sagittarius-like
dwarfs undergoing tidal strippings. A sample of 6,000 star particles is used for the dynamical constraint of each 
system, and the statistical uncertainties are smaller than or comparable to the symbol size. }
\label{fig:fit_vs_sf}
\end{figure*}

\begin{figure*} 
\begin{center}
\includegraphics[width=0.8\textwidth]{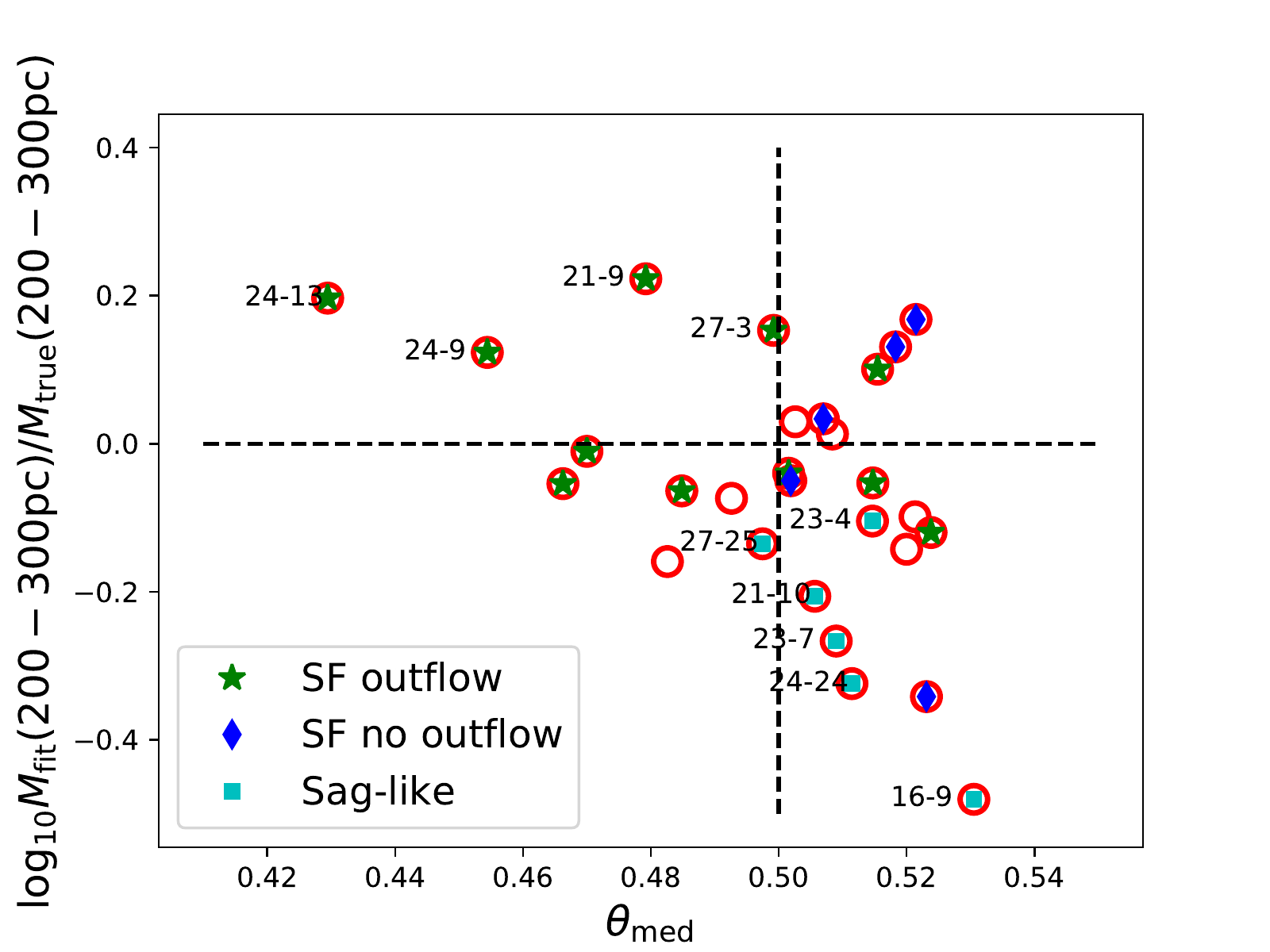}%
\end{center}
\caption{Differences between the best-fitting total mass between 200 and 300~pc and the truth in log space 
($y$-axis), reported as a function of the dynamical status of tracer star particles between 200 and 400~pc 
($x$-axis). The dynamical status is quantified by the medians of the radial action angle. A median of 0.5 
refers to the steady state. The larger the median deviating from 0.5, the more unrelaxed the system is. The 
adopted potential model to recover the underlying mass profiles, free parameters of the model and the meanings 
of filled symbols with different colors and shapes are the same as Figure~\ref{fig:fit_vs_angle}. The statistical
uncertainties in the best fits are smaller than or comparable to the symbol size. The horizontal and vertical 
black dashed lines mark zero bias in best fits and perfect steady state ($\theta_\mathrm{med}=0.5$), respectively.
Systems with under-estimated $M(200-300\mathrm{pc})$ are more likely to have the medians of their radial action
angle distribution biased to be larger than 0.5, and vise versa.}
\label{fig:fit_vs_angle}
\end{figure*}

\begin{figure*} 
\begin{center}
\includegraphics[width=0.8\textwidth]{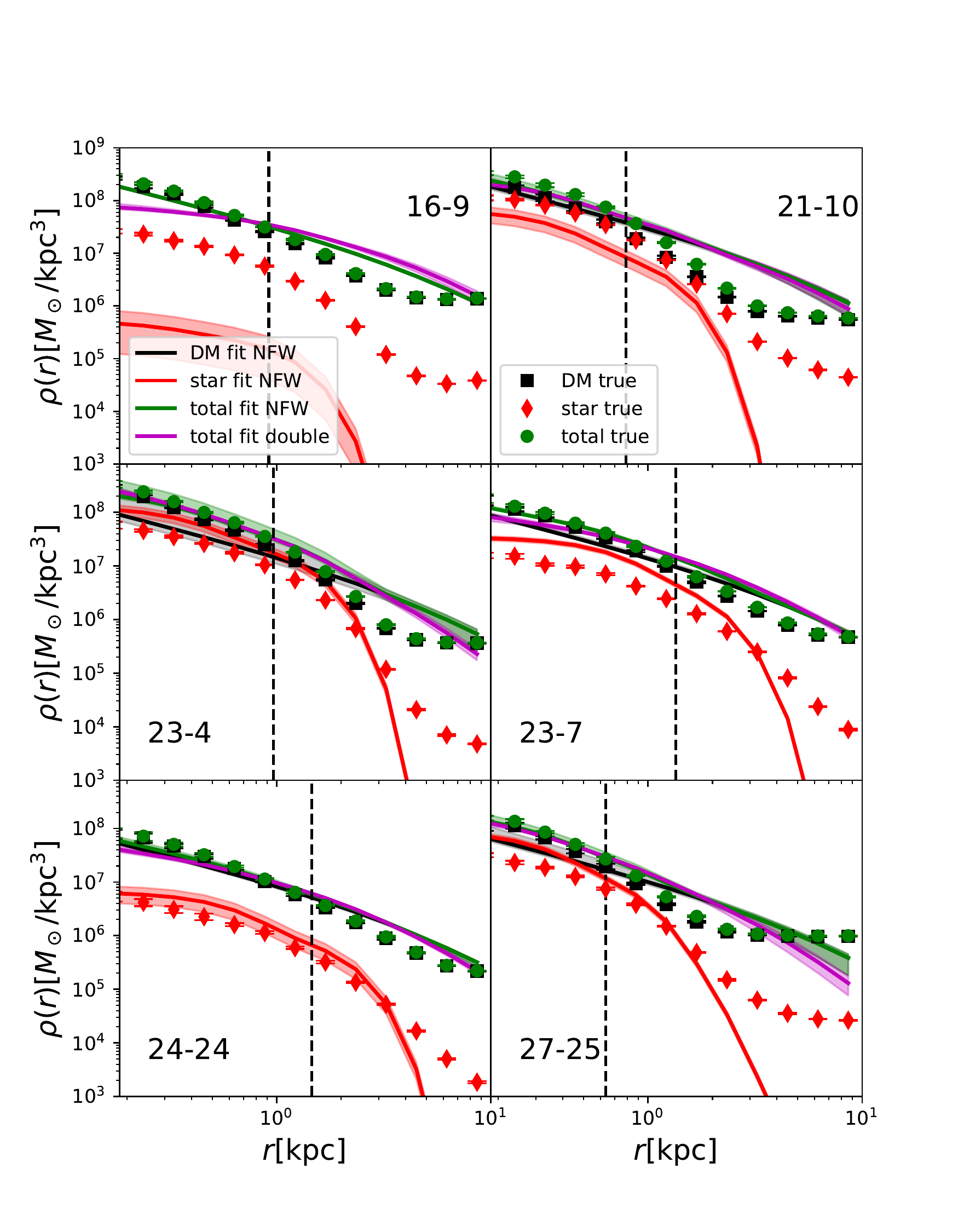}%
\end{center}
\caption{Black squares, red diamonds and green dots show the dark matter, star and total density profiles, 
calculated in spherical shells centered on six Sagittarius dSph-like systems. Here all particles at 
the corresponding radial ranges are used to calculate the true profiles. Solid curves with corresponding 
colors show the best-fitting profiles by \textsc{jam}, based on the NFW model profile. Notably, red 
solid curves are based on deprojections of the 2-dimensional MGEs, which are created with stars bound to 
the dwarfs. The magenta solid curves are best-fitting total profiles based on the double power law model profile,
with the inner slopes allowed to free, but the outer slopes fixed to 3. For simplicity, the stellar-mass-to-light 
ratios ($M/L$) of the luminous component are also fixed to unity when we use the double power law model. Shaded 
regions show the 1-$\sigma$ uncertainties of the best fits, while the errorbars of the true profiles are the 
1-$\sigma$ uncertainties of 100 boot-strapped subsamples of all particles. Only bound star particles are used as 
tracers, without the inclusion of observational errors. The gas and black hole components are significantly 
subdominant and are thus not shown. The $x$-axis range is chosen to be larger than 0.2~kpc, to avoid the very 
central regions which are affected by the resolution. Black vertical lines mark the half-mass radii. Numbers in 
each panel give the corresponding host system number and the satellite ID.
}
\label{fig:prof_fit}
\end{figure*}

\begin{figure*} 
\begin{center}
\includegraphics[width=0.8\textwidth]{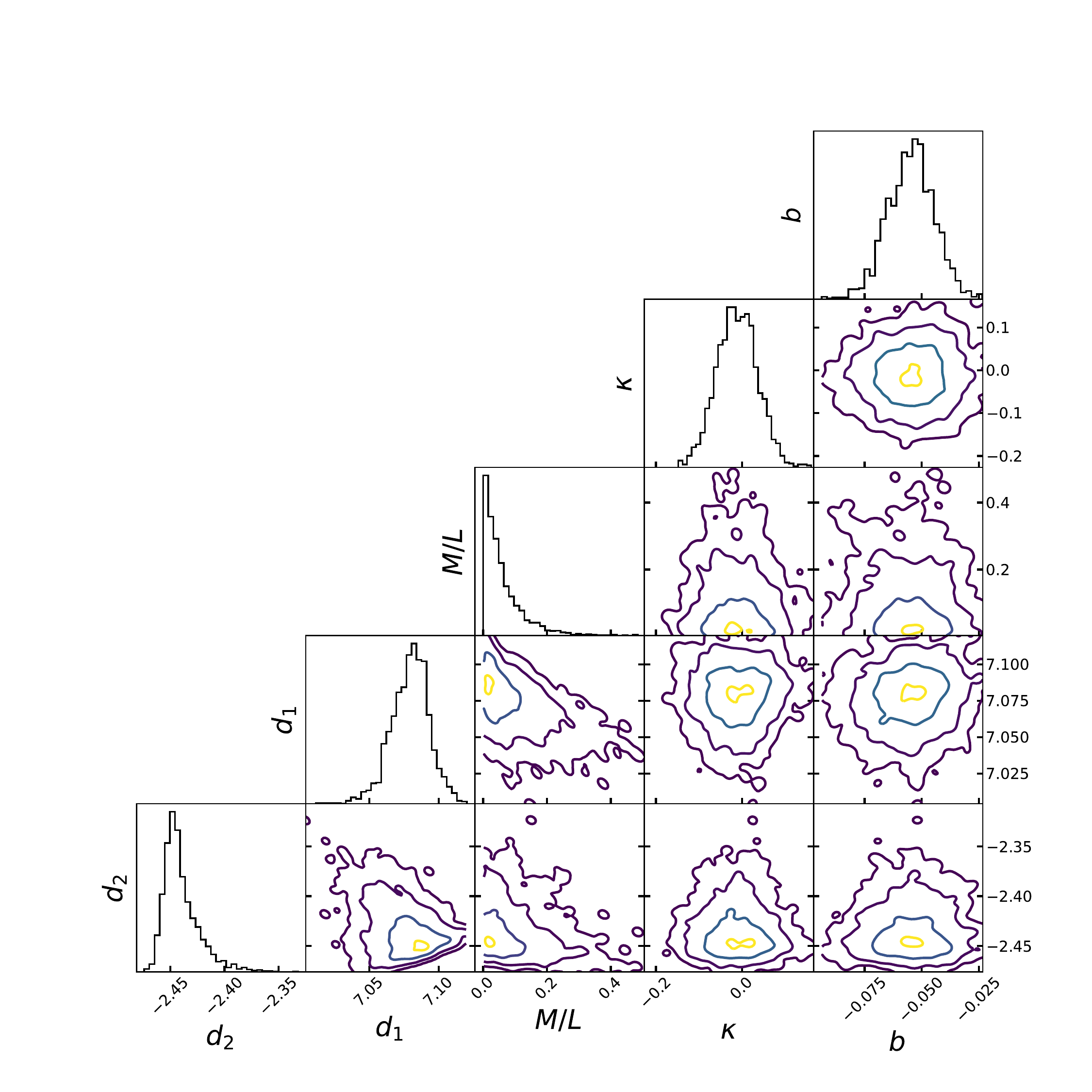}%
\end{center}
\caption{The likelihood contours of all possible combinations of every two out of the five model parameters for 
Au16-9 (NFW potential model with free $M/L$ and 6,000 star particles as tracers). The yellow contour includes 
10\% of the MCMC sample centered on the best fits. The light blue and two dark blue contours represent the 1, 2 
and 3-$\sigma$ confidence levels. The degeneracies between the tracer parameters ($\kappa$ and $b$) and the potential
parameters ($M/L$, $d_1=\log_{10}\rho_s^2 r_s^3$, and $d_2=\log_{10}\rho_s$) are very weak, but there are stronger degeneracies among $M/L$, $d_1$ and $d_2$.
}
\label{fig:contour}
\end{figure*}

The left plot of Figure~\ref{fig:fit_vs_sf} shows the best-fitting versus true masses for all 28 dwarfs. We 
show this for the mass enclosed within the half-mass radius of tracers, $M(<r_\mathrm{half})$, and the mass between
200 and 300~pc, $M(200-300\mathrm{pc})$. The fitting is based on the double power law model profile for the underlying 
dark matter (Equation~\ref{eqn:double}) with free inner slopes. Note the matter density within 150~pc is often used 
as the proxy to the inner slopes \citep[e.g.][]{2019MNRAS.484.1401R}. Unfortunately, 150~pc is below the resolution 
limit of \textsc{auriga}. So we focus on the radial ranges which are above the resolution limit for now, and we will
investigate the constraint within 150~pc later in Section~\ref{sec:corecusp}, subject to extrapolations 
to the very center. 

In general, $M(<r_\mathrm{half})$ is constrained better, with a scatter of $\sim$0.067~dex. This is in good agreement 
with the commonly accepted experience that the mass enclosed within the half-mass (or half-light) radius of tracers 
can be more robustly constrained upon dynamical modeling  \citep[e.g.][]{Wolf2010,2011ApJ...742...20W,2015MNRAS.453..377W,
2020SCPMA..6309801W}. This is due to the degeneracy between the two halo parameters, $\rho_s$ and $r_s$, reflecting 
that dynamical models mostly constrain the gravity or mass at the median radius of the tracers, but are less sensitive 
to the shape of the mass profile \citep[e.g.][]{2016MNRAS.456.1003H,2021MNRAS.505.3907L,2022arXiv220315268L}. There are 
a few cases for which $M(<r_\mathrm{half})$ are constrained slightly worse than $M(200-300\mathrm{pc})$, which are quite 
rare though.

$M(200-300\mathrm{pc})$ is recovered reasonably, with a scatter of 0.167~dex, which is larger than 
the scatter of $M(<r_\mathrm{half})$. We further classify systems into different groups according to 
their star formation activities. At first, we notice there are 11 star-forming systems with prominent 
galactic winds or gas outflows in the simulation, and on top of the corresponding red circles, we 
overplot green stars. An example of such outflows is shown in the left plot of Figure~\ref{fig:xykin_detect2}, 
which we will discuss later. There are another five systems with larger than zero star formation rates, but 
do not show obvious galactic winds, which we denote with blue diamonds. The six Sagittarius dSph-like (hereafter 
Sag-like) systems are demonstrated with cyan squares. Interestingly, systems classified in these groups show 
different trends. Quiescent Sag-like systems are more likely to have $M(200-300\mathrm{pc})$ under-estimated, 
whereas the best-fitting $M(200-300\mathrm{pc})$ for star-forming systems with strong outflows are biased to
to be more above the black dashed line by $\sim$0.05~dex on average. The other systems are more symmetrically
distributed on both sides of the black dashed line. 

In the right plot of Figure~\ref{fig:fit_vs_sf}, it is shown that the bias in best-fitting 
$M(200-300\mathrm{pc})$ from the truth correlates with the specific star formation rate (sSFR). 
The correlation is weak or absent for $M(<r_\mathrm{half})$. For the five systems with not zero 
star formation rates (SFR) but without prominent galactic winds, their sSFRs are weak, while the 
biases in their best fits are smaller on average. The six Sag-like systems are all quiescent with 
zero sSFR. Those systems with zero sSFR but are not too close to the galaxy center and are not 
yet undergoing strong tidal effects, i.e., not Sag-like, tend to show smaller biases in their 
best fits as well (red circles without star, diamond or square symbols). 

In order to understand the cause for such a correlation between the bias in best fits and the sSFR, we 
provide in Figure~\ref{fig:fit_vs_angle} the best-fitting versus true masses between 200 and 300~pc in 
log space, versus the dynamical status of the systems ($x$-axis) between 200 and 400~pc. In the $x$-axis 
of Figure~\ref{fig:fit_vs_angle}, $\theta$ is the radial action angle, or referred to as the phase angle
by \cite{2016MNRAS.456.1003H,2016MNRAS.456.1017H}. For a star at radius $r$, $\theta$ is defined as

\begin{equation}
    \theta(r)=\frac{1}{T}\int_{r_p}^{r}\frac{\ud r'}{v_r},
    \label{eqn:phase}
\end{equation}
where $T$ is the radial period, $r_p$ is the radius at pericenter and $v_r$ is the radial velocity. Note when 
radial cuts are adopted in the tracer sample, the integral boundaries have to be properly changed accordingly.
Equation~\ref{eqn:phase} can be easily evaluated for a spherical potential. Here we at first calculate potential 
profiles for each dwarf based on the true radial distribution of all particles (star, dark matter, gas and 
black hole) under the spherical approximation, and we evaluate $\theta$ for each tracer star particle using 
the potential profile. Since our systems are not strongly deviating from spherical symmetry, the spherical 
assumption should approximately hold for most of our dwarfs, though not all. For a system in steady state, 
$\theta$ evaluated from the true potential is expected to distribute uniformly between 0 and 1, with the 
median value close to 0.5. The deviation of the medians from 0.5 thus indicates the amount of deviations 
from steady states. This is why the $x$-axis quantity of Figure~\ref{fig:fit_vs_angle} is chosen as the 
median value of the $\theta$ for each system, which we denote as $\theta_\mathrm{med}$. The larger 
$\theta_\mathrm{med}$ deviates from 0.5 (the black vertical dashed line), the stronger the system is 
deviating from the steady state. 

We adopt particles between 200 and 400~pc to calculate the phase angles. The bias of best-fitting versus true 
$M(200-300\mathrm{pc})$ correlates with the dynamical status over this range\footnote{The choice of 200 to 
400~pc is empirical. We choose a slightly larger outer radius of 400~pc than the outer radius for $M(200-300\mathrm{pc})$,
i.e., 300~pc. This is because the orbits of tracers currently at larger radii can still extend to smaller 
radial range and affect the constraints there. However, we have checked other choices of radial ranges, 
such as 200 to 300~pc, 200 to 500~pc and 200~pc to 1~kpc, which all lead to some correlations between 
the bias in best fits and the corresponding dynamical status as well. Radial ranges larger than 1~kpc do not 
show any obvious correlation, which is partly because they do not fully represent the dynamical status in 
inner regions, and the recovery of the total mass between 200 to 300~pc mainly depends on tracers in more 
inner regions.}. The scatter is large, but there exists the trend that systems more strongly deviating 
from steady states between 200 and 400~pc are on average more likely to have larger biases in $M(200-300\mathrm{pc})$. 
Explicitly, we find systems with under-estimated $M(200-300\mathrm{pc})$ mostly tend to have $\theta_\mathrm{med}$ 
larger than 0.5. In other words, their phase angle distribution is biased to large values. On the contrary, 
many systems with over-estimated $M(200-300\mathrm{pc})$ tend to have $\theta_\mathrm{med}$ smaller than 0.5, 
i.e., phase angle distribution is biased to small values, though some of them still have $\theta_\mathrm{med}>\sim0.5$. 
Note for these systems with over-estimated $M(200-300\mathrm{pc})$ and $\theta_\mathrm{med}>0.5$, we find part 
of them look a bit flattened, such as the blue diamond at the upper right corner (Au27-2), so the calculation of 
their $\theta_\mathrm{med}$ might be slightly affected by the fact that we adopted spherical potential profiles to 
calculate their phase angles, but the overall trend is reasonable. 

Nevertheless, Figure~\ref{fig:fit_vs_angle} unambiguously indicates that the amount of bias in best fits is related 
to the dynamical status of the systems, and thus the correlation between the bias in best fits and the sSFR reflects 
the dynamical status behind\footnote{The deviation of $\theta_\mathrm{med}$ can be caused by either a) $\theta$ is 
evaluated with the correct potential, but tracers or the system are not in steady states; b) $\theta$ is evaluated 
for steady state tracers but with the wrong potential. In our case, it is the former.}. For systems with strong 
galactic winds, despite the fact that wind particles are not used as tracers in our analysis, the galactic winds 
have caused deviations from steady states, resulting in larger biases (mostly over-estimates in
$M(200-300\mathrm{pc})$ in their best fits). For the other five systems with larger than zero but weak star formations,
perhaps because they do not present prominent outflows to disturb the system, they tend to be closer to steady states 
on average and also have smaller biases in the best fits. 

On the other hand and very interestingly, we find many of the Sag-like systems are undergoing some level of 
contractions along the longer image axis\footnote{For Au16-9, it is the $y'$-axis, while for the other systems, 
it is the $x'$-axis. The $x'$ or major axis of the image plane, which is defined through the spin direction, 
usually corresponds to the longer axis for systems which are not strongly triaxial. However, due to the 
mis-alignment between the spin axis and the minor axis of the dwarf systems (see Section~\ref{sec:star}), 
sometimes the image major/$x'$ axis is slightly shorter than the minor/$y'$ axis. When it happens, the difference 
between $x'$ and $y'$ axes is usually very small, such as Au16-9. If the difference is big, the systems are 
strongly triaxial and are thus excluded from our sample.}, which perhaps result in the under-estimated
$M(200-300\mathrm{pc})$. These include Au16-9, Au23-7 and Au24-24, as marked in the figure with corresponding 
numbers. An example of the contraction is shown in the left panel of Figure~\ref{fig:xykin_detect} for Au16-9, 
which we will discuss later. In addition to the three systems, the other three Sag-like systems in Table~\ref{tbl:6sag} 
(Au21-10, Au23-4 and Au27-25) show less significant underestimates in $M(200-300\mathrm{pc})$. We only see some 
very weak contractions along the $x'$ axis for Au21-10 and Au23-4, and this is perhaps why the best fits for the 
two dwarfs agree better with the truth. Au27-25 undergoes strong expansions along the $y'$ and $z'$ axes, 
and along the $x'$ axis, it shows contractions. Note the expansion is not due to galactic winds as all the 
Sag-like systems have zero SFR and no wind particles. 

The contractions and expansions in these Sag-like systems are due to the strong tidal effects at small distances. 
In fact, all six Sag-like systems are undergoing some expansions along the line of sight. This is expected, 
because the tidal disruption happens along the radial direction pointing to the galaxy center, and although 
the artificial observer is 8~kpc away from the galaxy center, the line-of-sight direction still aligns with 
the radial direction. As shown by \cite{2022MNRAS.510.2724O}, when the dwarf is close to its peri-centric 
passage, the effective radius first undergoes a small decrease, followed by a large increase, sometimes 
associated with oscillations. On the contrary, if without tidal forces, \cite{2022MNRAS.510.2724O} showed 
that the effective radius stays almost constant. 

Our results seem to indicate, strong tidal effects, which can cause both contractions and expansions, 
tend to cause underestimates in $M(200-300\mathrm{pc})$, and perhaps it is the contraction that is more 
related to the under-estimated $M(200-300\mathrm{pc})$. Perhaps the coherent contractions have 
caused smaller velocity dispersions in inner regions and hence under-estimated inner profiles. Besides, 
due to the global contraction, the apocenters of tracer orbits likely move inwards with time, resulting 
in $\theta_\mathrm{med}$ greater than 0.5 if $\theta$ is evaluated based on the current underlying potential. 
This is because, the potential changes with time, but the apocenters will be over-estimated if they are
calculated from the current potential by assuming that the potential does not evolve (steady state). On 
the other hand, outwards motions or expansions, such as the gas outflows or galactic winds, are more likely 
to cause overestimates in $M(200-300\mathrm{pc})$. The galactic winds cause lowered density in central 
regions, driving the whole system out of equilibrium. When $\theta$ is evaluated based on the current 
potential, assuming the system is in steady state, the distribution of $\theta$ is biased to be smaller,
with $\theta_\mathrm{med}$ less than 0.5.

In the next subsection, we move on to perform more detailed investigations on the best fits for the six 
quiescent Sag-like systems and another four star-forming systems with strong outflows.

\subsection{Examples of the best-fitting matter density profiles}
\label{sec:profile}

\begin{figure} 
\includegraphics[width=0.49\textwidth]{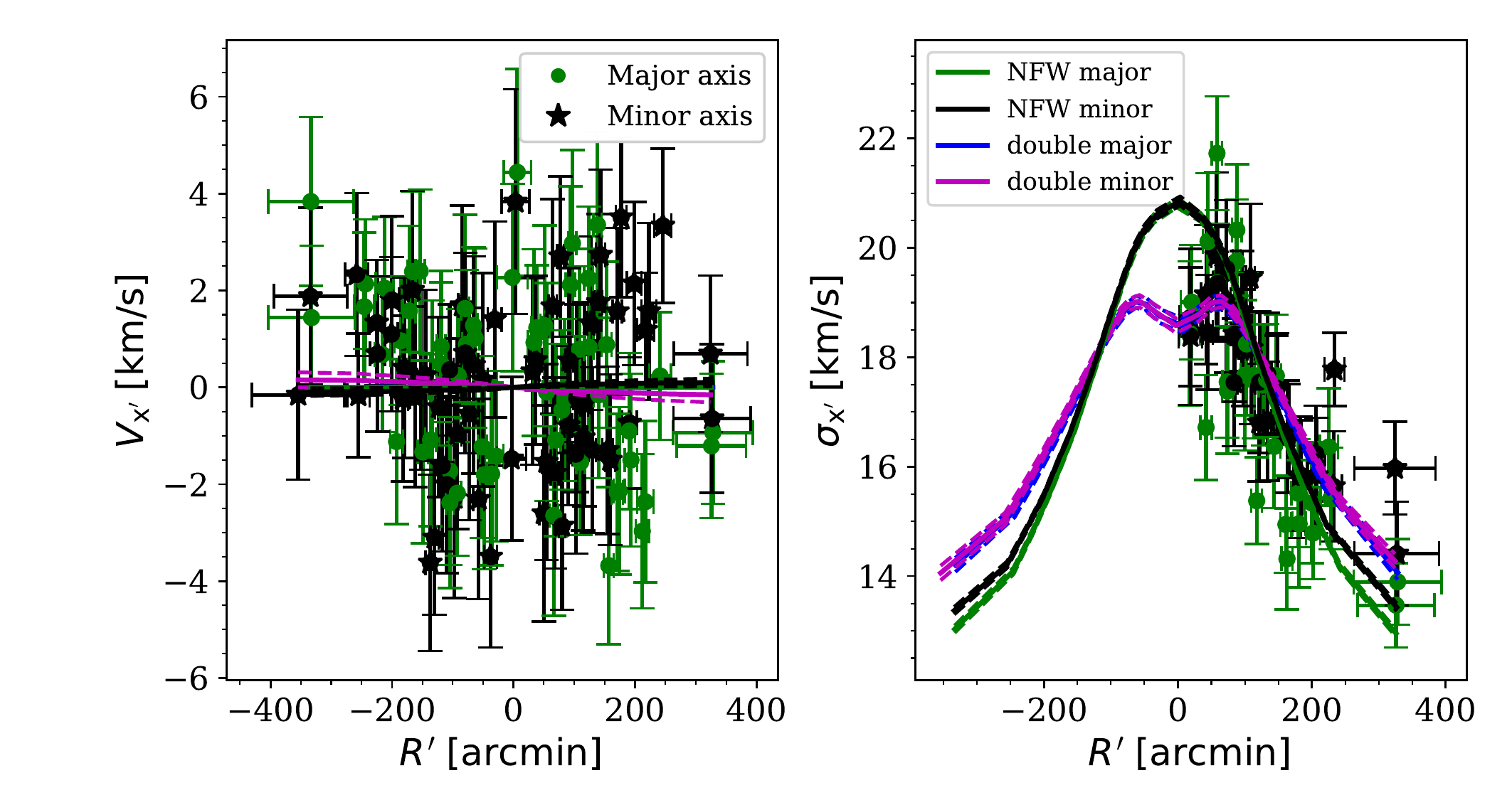}%
\caption{The $x'$ component of the mean velocity (left) and velocity dispersion (right) profiles of star particles, 
binned along the major (green) and minor (black) axes for Au16-9. Note along the major and minor axes, only stars 
within sectors of $\pm45$ degrees to the corresponding axes are used to calculate the mean velocity and velocity dispersion 
in each bin. Each bin contains 100 stars. The $x$ and $y$ errors indicate the bin width and the 1-$\sigma$ scatters, 
respectively. Green and black solid curves show model predictions along the major and minor axes, based on the NFW model 
profile. Blue and magenta solid curves show the model predictions along the major and minor axes, based on the double power 
law model profile for the dark matter with free inner slopes. Dashed curves around the solid ones with corresponding colors 
indicate the 1-$\sigma$ uncertainties in the model. The green, black, blue and magenta curves in the left panel are all very 
close to horizontal lines at $y=0$, and hence are overlapping with each other. }
\label{fig:xykin_nfw}
\end{figure}

\begin{figure*} 
\begin{center}
\includegraphics[width=0.8\textwidth]{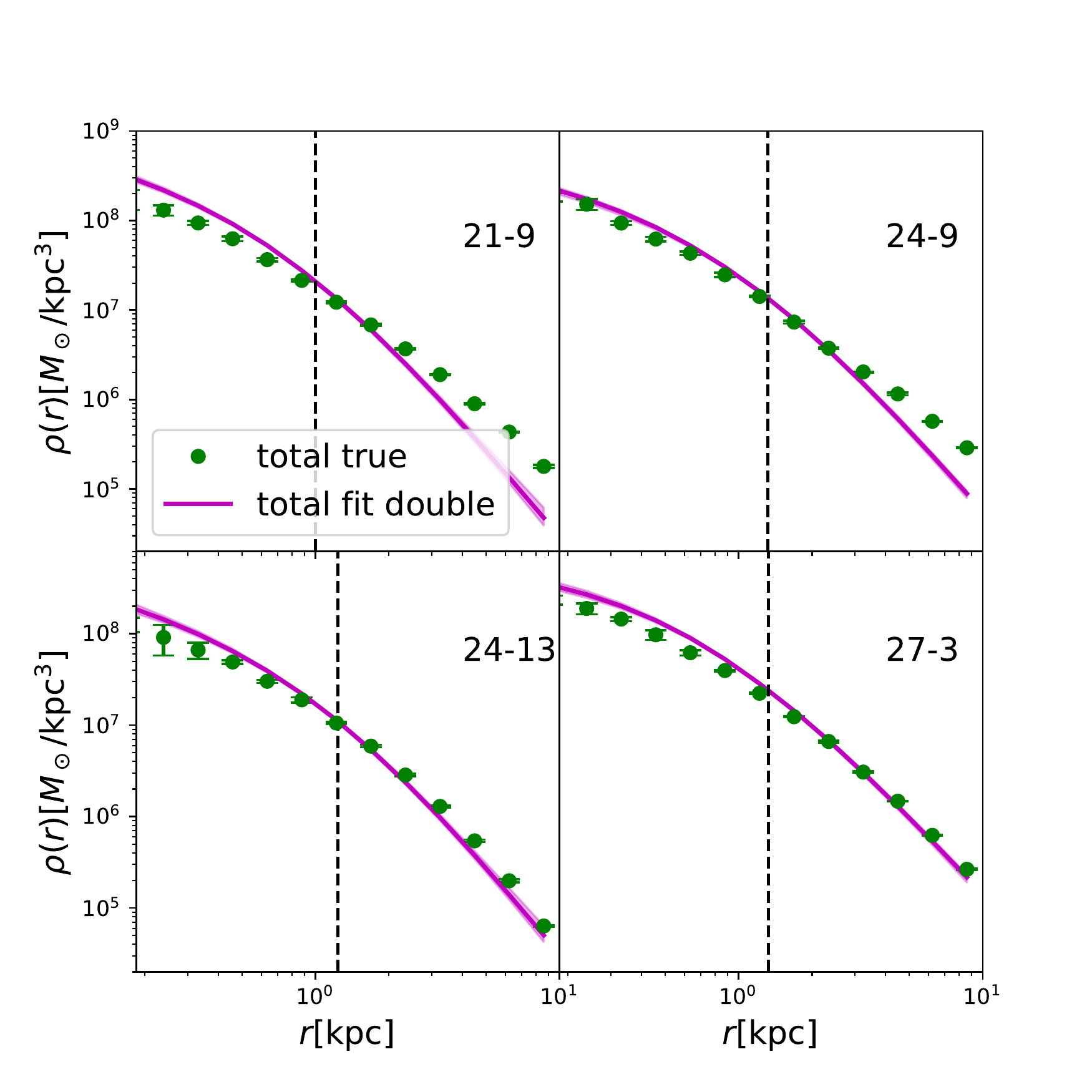}%
\end{center}
\caption{Similar to Figure~\ref{fig:prof_fit}, but shows the true (green dots) and best-fitting (magenta curves)
total mass profiles for four representative star-forming systems with outflows. Black vertical lines mark the 
half-mass radii.
}
\label{fig:prof_fit_sf}
\end{figure*}

\begin{figure} 
\includegraphics[width=0.49\textwidth]{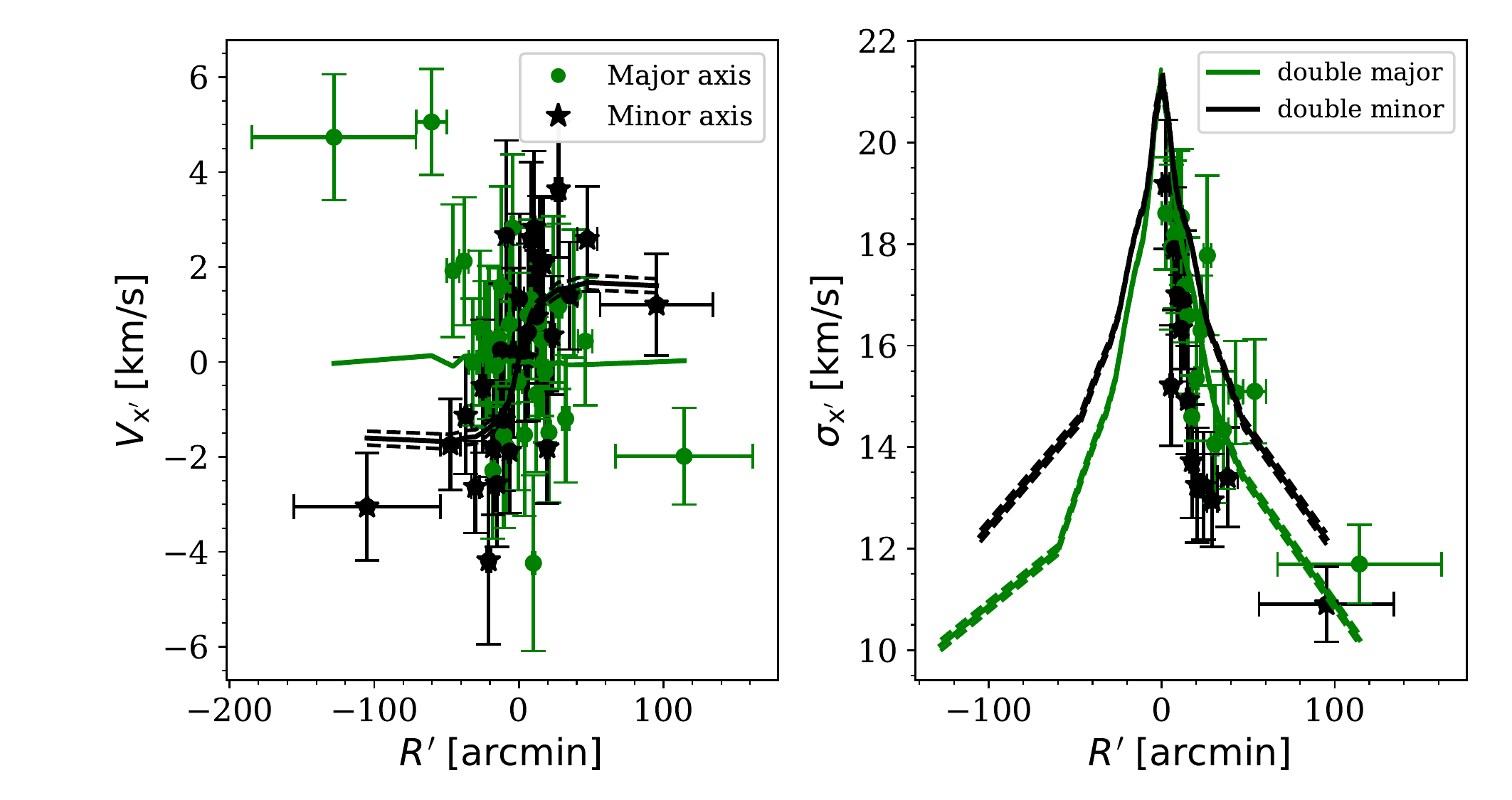}%
\caption{Similar to Figure~\ref{fig:xykin_nfw}, it shows the $x'$ component of the mean velocity (left) and 
velocity dispersion (right) profiles, of star particles binned along the major (green) and minor (black) axes 
for Au21-9.}
\label{fig:xykin_sf}
\end{figure}

In this subsection, we further investigate the best-fitting and true density profiles for six Sag-like 
systems and other four systems with prominent galactic winds. 

\subsubsection{Best-fitting density profiles and velocity maps for Sagittarius dSph-like systems}

We start with the six Sag-like systems. In Figures~\ref{fig:fit_vs_sf} and \ref{fig:fit_vs_angle}, we have 
shown that the best-fitting masses between 200 and 300~pc, $M(200-300\mathrm{pc})$, are under-estimated for  
the six systems and to different levels, when using the double power law model profile for the underlying dark 
matter. In Figure~\ref{fig:prof_fit}, the best fits (magenta curves) based on the double power law potential 
model and true total mass profiles (green dots with errorbars) are shown for the six systems. For Au16-9, 
Au21-10, Au23-7 and Au24-24, their best-fitting inner profiles are lower than the truth, i.e., the magenta 
solid curves are below the green dots within $\sim$0.5~kpc, and the differences are significant compared with 
the small errorbars. $M(200-300\mathrm{pc})$ of Au23-4 and Au27-25 are mildly under-estimated, so the 
differences between the magenta curves and the green dots are not very obvious in the corresponding panels. 

$M/L$ is fixed to unity for the magenta curves, but the underestimates in the inner profiles 
remain the same, even if we allow $M/L$ to be free. Besides, we have fixed the outer slopes to be 
$\alpha=3$. However, the poor constraints in the outskirts are not improved if we allow the outer 
slopes to be free. This is probably due to the significantly smaller number of tracer star particles 
in outskirts. These systems mostly have over-estimated outer profiles and under-estimated inner 
profiles. The true and best-fitting profiles cross with each other at the radii 
close to the half-mass radii. Note although only bound particles are used as tracers, the true 
profiles in Figure~\ref{fig:prof_fit} are calculated based on all particles at the corresponding radial 
ranges\footnote{Inclusion of unbound particles mainly affects the true profiles in outskirts, with the 
inner profiles barely affected. }. This is because though particles belonging to the host halo 
might be approximated to have a constant density over the scale of the dwarf, they still contribute to 
the potential gradient. Since background particles belonging to the host halo could be more dominant 
for such nearby Sag-like systems, we have also tried to model the underlying matter distribution using 
the double power law model plus a constant density model. However, the best constrained model still 
tends to under-estimate the inner profiles and over-estimate the outer profiles.

We also show in Figure~\ref{fig:prof_fit} the best-fitting star and dark matter profiles separately. The 
red, black and green curves are the best-fitting star, dark matter and total profiles based on the NFW model 
profile. Here $M/L$ is allowed to be free. It is clearly shown that the profiles of the stellar component, 
and hence $M/L$, are poorly constrained by \textsc{jam}, with the best fits (red curves) significantly deviating 
from the truths (red diamonds) in most cases. The dark matter profiles, on the other hand, can be constrained 
better, but still show some significant deviations from the truth in inner regions, especially for Au23-4 and 
Au23-7. Fortunately, despite the relatively worse fits to the stellar and dark matter components individually, 
the total profiles can be fit significantly better.

Due to the poorly constrained $M/L$, we have fixed $M/L$ to unity in Figures~\ref{fig:fit_vs_sf} and \ref{fig:fit_vs_angle}, 
when the double power law potential model is used. However, with the double power law model, we do have tried 
to allow $M/L$ to be free as well, though we choose not to explicitly show the results to avoid redundancy. Our 
conclusions remain very similar. The stellar and dark matter profiles are poorly constrained individually. Actually, 
with the free inner slopes in the double power law model, the fitting to the dark matter component can become 
significantly lower than the truth in inner regions for a few cases, whereas the stellar component is significantly 
over-estimated, with the total profiles (stellar $+$ dark matter) still more reasonably constrained. Hereafter, unless 
otherwise specified, $M/L$ will be fixed to unity, i.e., its true value. This is a reasonable operation, because 
observationally, $M/L$ can be alternatively constrained through stellar population synthesis modeling and fixed 
upon dynamical modeling.

The poor fits to stellar and dark matter profiles individually are mainly due to the degeneracy between 
$M/L$ and halo parameters. First of all, the true stellar and dark matter profiles share similar shapes in
Figure~\ref{fig:prof_fit}. Note all particles have been included in these true profiles, and if only using 
bound particles, the stellar and dark matter profiles would appear more similar in outskirts. This is due to 
the strong tidal stripping happened to these Sag-like systems, making them have similar outer dark matter and 
stellar halo profiles \citep[e.g.][]{2022MNRAS.511.6001E}. As a result, the model has difficulties distinguishing 
the stellar and dark matter components, as they have similar contributions to the shape of the underlying potential
profiles. This explains why the best fits to stellar and dark matter components are poor individually. Moreover, 
we show in Figure~\ref{fig:contour} the likelihood contours of different parameter combinations for Au16-9. The
anti-correlations or degeneracies among $M/L$ and the two halo parameters ($d_1$ and $d_2$) are prominent.

Comparing the magenta and green curves in Figure~\ref{fig:prof_fit}, we find the inner profiles are constrained 
worse when the double power law model profile is used, for Au16-9, Au21-10, Au23-7 and Au24-24, than the NFW profile. 
To explore why the inner profiles are constrained worse when the inner slopes are allowed to free, we check the 
best-fitting and true velocity maps in Figure~\ref{fig:xykin_nfw}, which shows the mean velocity and velocity 
dispersion profiles for the $x'$ component and for Au16-9 as an example. Both the velocity and velocity dispersion 
profiles have large scatters, and the double power law model tends to go through the middle of the velocity 
dispersion profiles in inner regions, while the NFW model leads to a curve which is higher in the center. Both 
are reasonable fits, but in fact the double power law model achieves a better fit with log likelihood larger than 
that of the NFW model by $\sim$33, which is significantly larger than the expected 1-$\sigma$ uncertainty of five 
free parameters. This is in fact benefited from our usage of a large sample of 6,000 member stars. While the 
velocity profiles are better fit with the double power law model profile, the deviation of the best-fitting 
double power law model from the true density profile is larger, which reflects the deviation from the steady 
state due to strong tidal effects, and thus a better fit to the velocity maps lead to a worse fit to the density 
profile. Explicitly, as we have mentioned, the strong contraction of member stars in Au16-9 (see 
Figure~\ref{fig:xykin_detect}) has likely caused smaller velocity dispersions in inner regions, and thus the 
double power law model with under-estimated inner profiles fits better the velocity map. By using the NFW model, 
more masses are forced to be in the inner region, leading to a better match to the true density profile but a 
worse fit to the velocity map. 

\subsubsection{Best-fitting density profiles and velocity maps for star-forming systems with outflows}

Now we move on to discuss a few examples of star-forming systems with prominent outflows.
Figure~\ref{fig:prof_fit_sf} shows the true and best-fitting total profiles for four representative star-forming 
systems. They all have strong outflows and large biases in their best-fitting $M_\mathrm{fit}(200-300\mathrm{pc})$. 
The four systems are marked with corresponding numbers in Figure~\ref{fig:fit_vs_angle}. For simplicity, we do not 
show the profiles for the stellar and dark matter components, but only show the total profiles. These systems all 
have over-estimated inner density profiles and under-estimated outer profiles, with the true and best-fitting 
profiles cross at roughly the half-mass radii. 

The mean velocity and velocity dispersion profiles of star particles for Au21-9 is shown in 
Figure~\ref{fig:xykin_sf}, and we only show the $x'$ component as an example. The apparent sizes of such 
distant star-forming dwarfs are smaller than nearby Sag-like objects in Figure~\ref{fig:xykin_nfw}. Overall, 
the agreement between the truth and the best fit is reasonable. The model tends to fit a rotation, 
i.e., $v_x'$ positive along the positive $y'$-axis (minor axis) and negative along the negative $y'$-axis. 
These systems have strong galactic winds (wind particles), but the star particles do not show obvious signatures 
of outwards motions. However, there are some indications of inflows of the stars on large scales along the 
major axis, that is, $v_x'$ is negative along the positive $x'$-axis (major axis) and positive along the 
negative $x'$-axis. Such inflows are perhaps due to tidal effects, because this dwarf is on its way of 
approaching host galaxy center. Besides, the stellar system can expand in response to the lowered potential
due to gas outflows at the beginning, which undergoes some overshooting and as a result would contract and 
get stabilized later \citep[e.g.][]{1996MNRAS.283L..72N,2022arXiv220607069L}. Maybe we happen to observe 
this dwarf at the contraction stage. The best-fitting velocity dispersion along the major axis is slightly 
lower than the truth beyond 40\arcmin, while the best-fitting dispersion along the minor axis seems to be a 
bit higher in amplitude than the truth beyond 10\arcmin, which are perhaps related to the deviation from 
steady states in outskirts. 

Comparing Figures~\ref{fig:prof_fit} and \ref{fig:prof_fit_sf}, it seems both figures have the best-fitting 
and true total profiles crossing at approximately the half-mass radii. As we have mentioned, due to the degeneracy 
between the halo parameters, most dynamical models mainly constrain the mass at the median radius of the tracers, 
but are less sensitive to the shape of the mass profile. To maintain the relatively robust constraints on the mass 
at the half-mass radius of tracers, the biases in the best-fitting inner and outer density profiles for the same 
system often happen in opposite ways. In addition, quiescent Sag-like systems and star-forming systems also tend 
to have opposite trends. Sag-like systems tend to have under-estimated inner profiles and over-estimated outer 
profiles, whereas star-forming systems with prominent outflows tend to have over-estimated inner profiles and 
under-estimated outer profiles. This is because, as we have explained in the previous subsection, the
contractions/expansions (perhaps mainly the contractions) due to strong tidal effects for nearby Sag-like 
systems and galactic winds/gas outflows for more distant star-forming systems drive the systems out of 
equilibrium, but in opposite ways. 

\subsection{Constraints based on less tracers and only line-of-sight velocities with observational errors}
\label{sec:err_vronly_2000}

\begin{figure} 
\includegraphics[width=0.49\textwidth]{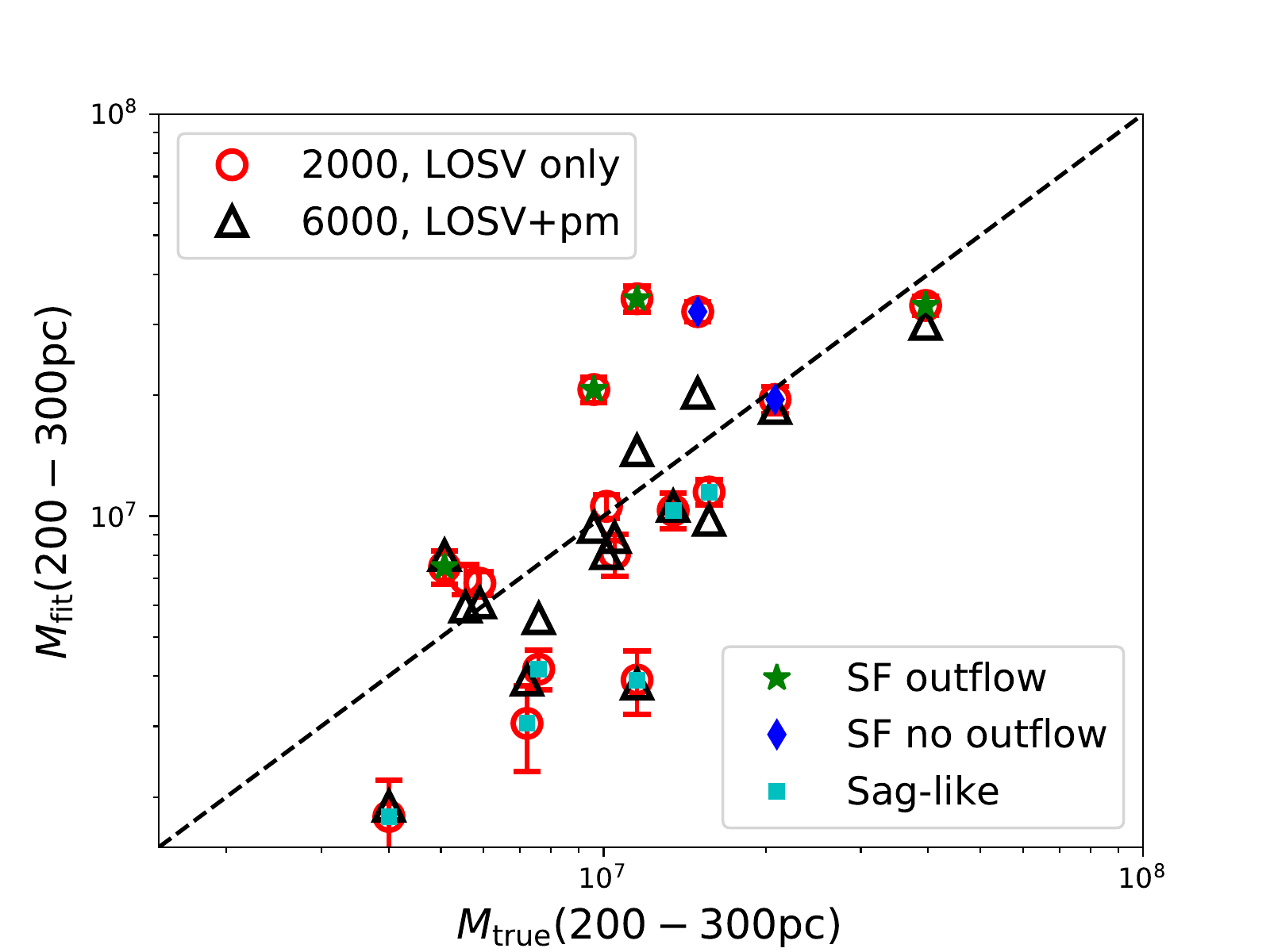}%
\caption{Red circles are besting-fitting versus true masses between 200 and 300~pc. It is similar to 
the left plot of Figure~\ref{fig:fit_vs_sf}, but we use a smaller tracer sample of 2,000 bound star particles, 
and only use their line-of-sight velocities after including observational errors (fore/background contamination 
is not modeled). After excluding systems with less than 2,000 bright stars (see Section~\ref{sec:err_bkgd} for 
details), only 16 systems left. The meanings of filled symbols with different colors and shapes have are the 
same as Figure~\ref{fig:fit_vs_sf}. Black empty triangles are repeats of the corresponding red empty circles 
in Figure~\ref{fig:fit_vs_sf}. Errors for back empty triangles are similar to or smaller than the symbol size, 
and thus are not shown. 
}
\label{fig:masstrue200_vs_massfit200_2000}
\end{figure}

Our results in the previous sections are based on a large sample of 6,000 star particles as tracers, 
with both line-of-sight velocities and proper motions, but without errors. The error-free and large 
sample of tracers have enabled us to control the size of statistical errors while focus on investigating 
the intrinsic systematics behind \textsc{jam}. However, current observations of nearby dwarf galaxies at 
most provide kinematical information of $\sim$2,000 member stars \citep[e.g.][]{2016MNRAS.463.1117Z}, 
most of which do not have accurate proper motion measurements. Thus for more practical meanings, in this 
subsection we repeat our analysis by using a smaller sample of 2,000 star particles as tracers, and we 
only use their line-of-sight velocities after including observational errors (see Section~\ref{sec:star} 
for details). After excluding distant systems which cannot have more than 2,000 stars brighter than $g=21$, 
we end up with 16 systems. 

The best constrained $M_\mathrm{fit}(200-300\mathrm{pc})$ versus their true values is shown in
Figure~\ref{fig:masstrue200_vs_massfit200_2000}. We overplot the error-free constraints based on 
6,000 star particles as black empty triangles. Approximately, it seems $M(200-300\mathrm{pc})$ can 
still be constrained ensemble unbiasedly, but compared with the measurements based on 6,000 star 
particles and both line-of-sight velocities and proper motions, the measurements now slightly biased 
more above the diagonal reference line at $M(200-300\mathrm{pc})>10^7\msun$. The scatter is increased 
by $\sim$50\%, with the black triangles having a scatter of 0.169~dex\footnote{This is slightly 
different from the scatter of red circles in the left plot of Figure~\ref{fig:fit_vs_sf}, due to the 
exclusion of distant systems with less than 2,000 bright stars. } and the red circles having a 
scatter of 0.260~dex. Our results thus indicate that systematic uncertainties caused by deviations from 
the steady states of the dwarf systems are dominating the final errors of mass estimation. If without 
accurate proper motion measurements, line-of-sight velocities of $\sim$2,000 tracers with typical 
observational errors can still achieve reasonable and approximately ensemble unbiased measurements 
for $M(200-300\mathrm{pc})$ on average, but with slightly larger scatters.

\subsection{Constraints on Sagittarius dSph-like systems after including observational errors and unbound star contamination}

\begin{figure*} 
\begin{center}
\includegraphics[width=0.8\textwidth]{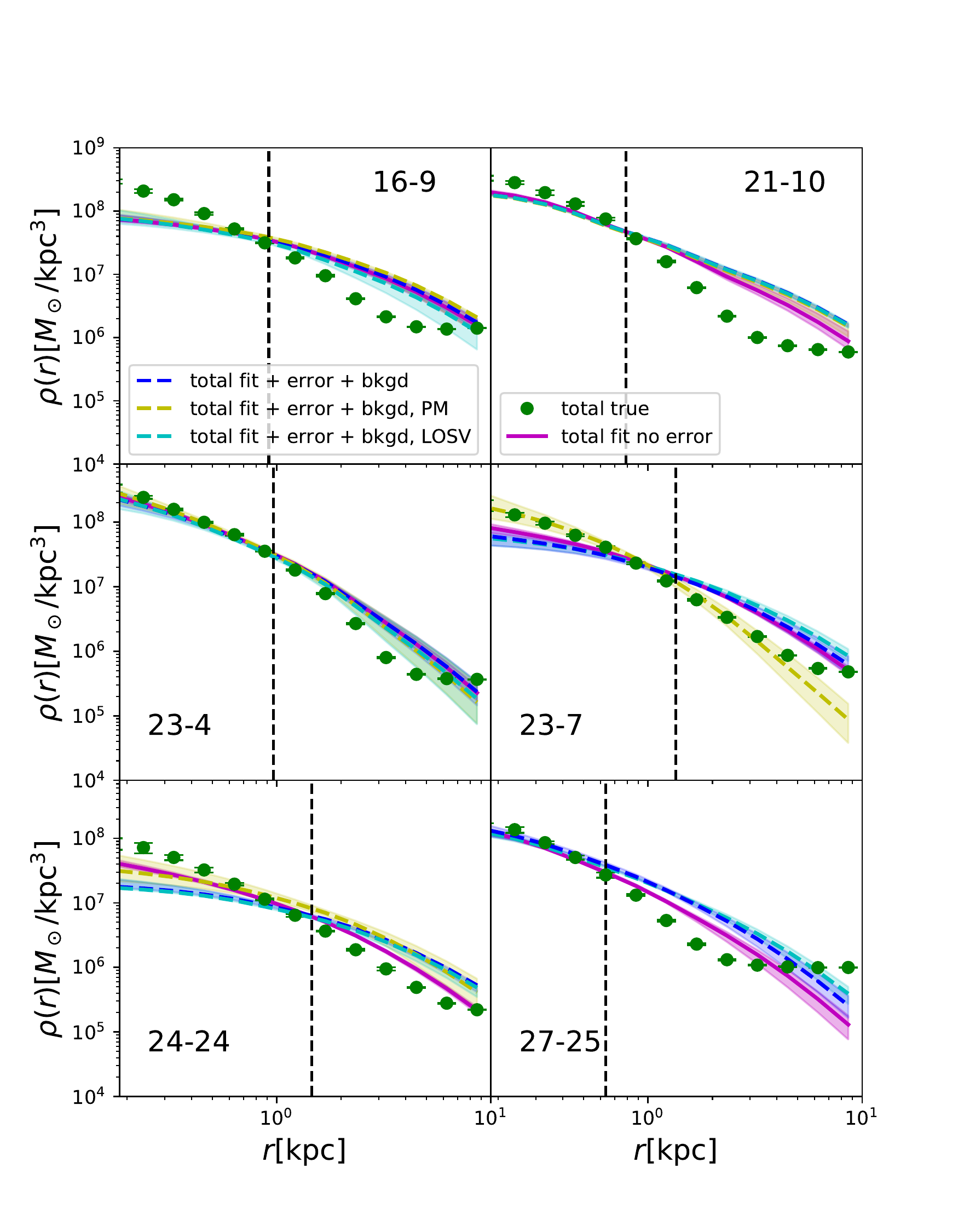}%
\end{center}
\caption{Similar to Figure~\ref{fig:prof_fit}. Green dots with errorbars are the true total density profiles 
of the six Sagittarius dSph-like systems, and magenta solid curves are the best fits based on the double 
power law functional form for the underlying potential (errors not modeled). These are exactly the same as 
in Figure~\ref{fig:prof_fit}. We further include realistic observational errors ($1-10$~km/s of errors in
line-of-sight velocities and {\it Gaia} DR3 proper motion and parallax errors), and the tracer stars are
selected according to their difference in kinematics with respect to the dwarf centers (see
Section~\ref{sec:err_bkgd} for details). Blue dashed curves are best fits based on tracer star particles
selected after including the observational errors. Cyan and yellow dashed curves are best fits based on 
only line-of-sight velocities and on only proper motions, respectively, with observational errors included. 
}
\label{fig:prof_infree_ml1_err_bkgd}
\end{figure*}

\begin{figure} 
\includegraphics[width=0.49\textwidth]{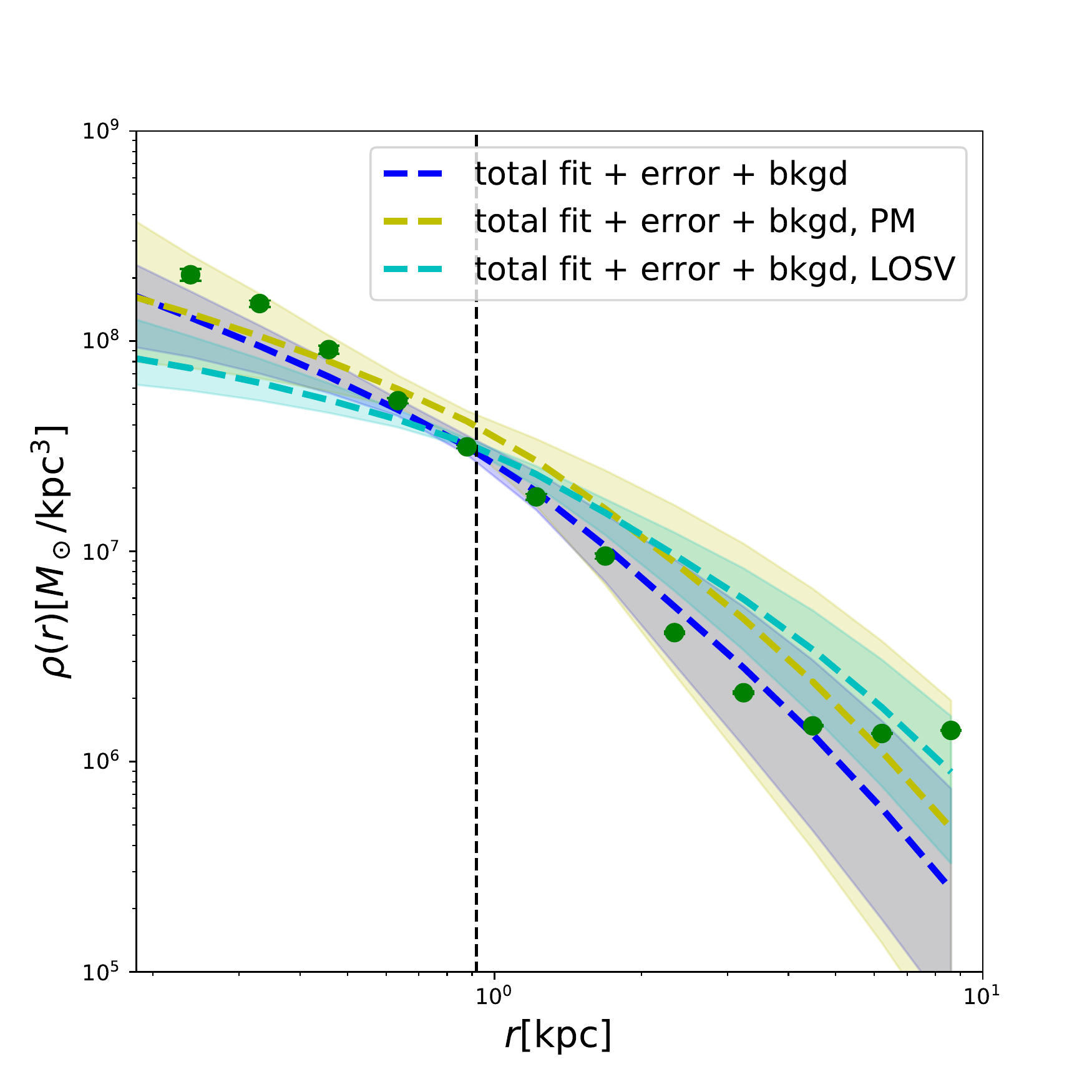}%
\caption{Similar to Figure~\ref{fig:prof_infree_ml1_err_bkgd}, but shows the results after incorporating CSST 
proper motion errors for only Au16-9. 
}
\label{fig:prof_infree_ml1_err_bkgd_csst}
\end{figure}

In the previous subsections, our results are based on bound star particles as tracers. In this section we discuss 
the results after incorporating line-of-sight velocity errors, {\it Gaia} DR3 parallax errors and either 
{\it Gaia} DR3 or CSST \citep[CSST;][]{zhan2011,Cao2018,Gong2019} proper motions errors. We also select member 
stars according to their kinematics. We first present results based on {\it Gaia} DR3 errors for the six Sag-like
systems. We then move on to show results based on the CSST proper motion errors for the dwarf with the closest 
distance to the mock observer (Au16-9). 

After including typical line-of-sight velocity errors, {\it Gaia} DR3 proper motions and parallax errors, we 
select tracer star particles based on their differences in proper motions, line-of-sight velocities and parallaxes 
with respect to the dwarf centers (see Section~\ref{sec:err_bkgd} for details), and the best fits are shown as 
blue dashed curves in Figure~\ref{fig:prof_infree_ml1_err_bkgd} for the six Sag-like systems. Stars selected 
in this way not only include those true bound member stars, but may also include fore/background contamination 
and some unbound star particles at the vicinity of the dwarf. As we have explicitly checked, because the associated 
observational errors are relatively small for such nearby Sag-like systems, the fraction of actual foreground or 
background stars is very small. In addition to bound member stars, most of the other tracer stars selected in 
this way are those unbound star particles surrounding the dwarf system. Part of these star particles might once 
be bound to the dwarf, and have been stripped by the host halo recently. 

After including observational errors and contamination by unbound stars, the uncertainties as indicated by the 
blue shaded regions are now significantly larger than the original magenta shaded regions, but the best fits stay 
very similar to the original error-free best fits. On the basis of the blue dashed curves, yellow and cyan dashed
curves are best fits using only proper motion data or using only line-of-sight velocities, respectively. For Au16-9, 
Au21-10, Au23-4 Au24-24 and Au27-25, results based on pure proper motions and pure line-of-sight velocities are 
consistent with each other within 1-$\sigma$. However, for Au23-7, pure proper motions and pure line-of-sight 
velocities show more significant differences. Interestingly, the yellow curves, i.e., with pure proper motions, 
show much better agreement with the true profiles. The larger difference between the yellow and cyan curves for 
Au23-7 is mainly related to how the tracer stars are selected. We found those unbound star particles selected as 
tracers and mainly in outskirts for this systems show larger velocity dispersions in their line-of-sight velocities 
than those of bound particles. The line-of-sight velocity distribution of these particles also deviates from the 
assumed Gaussian distribution by \textsc{jam}, and \textsc{jam} does not model them as background. As a result, 
the increased velocity dispersions lead to the much higher amplitudes in outskirts of the best fits, when both 
line-of-sight velocities and proper motions are used and when only line-of-sight velocities are used, but is not 
the case when only proper motions are used.

Our results indicate that, for such nearby Sag-like systems, we can observe enough number of bright 
stars, and with the accuracy of {\it Gaia} DR3 errors, proper motions can be as useful as or even 
sometimes better than line-of-sight velocities. For example, for Au16-9 at slightly beyond 20~kpc, 
the typical {\it Gaia} DR3 proper motion error of 0.1~mas/yr corresponds to error in tangential velocity 
of $\sim$5~km/s, which is comparable to the typical error of 1 to 10~km/s for the line-of-sight velocity. 
We have repeated our analysis by incorporating observational errors for more distant dwarf galaxies. 
Unfortunately, for systems close to or beyond 100~kpc, the number of bright stars is limited, 
and {\it Gaia} DR3 proper motion errors are not good enough to achieve reasonable constraints, with 
the fitting based on pure proper motion data very difficult to converge. This is expected, because 
most stars with $18<g<19$ have typical {\it Gaia} DR3 errors of 0.1~mas/yr, which corresponds to 
$\sim$40~km/s of errors in tangential velocities at $\sim$80~kpc, and thus are useless.

However, the precision in future {\it Gaia} 
proper motion data is expected to be improved in a way proportional to $t^{1.5}$ ($t$ is the time), 
indicating more than a factor of two reduction in error for DR4 compared with DR3, and additionally a 
factor of $\sim$2.8 of improvement for the extended mission compared with the end of the 5-year mission 
at fixed magnitude, pushing to a limit which can even be better than the typical line-of-sight velocity 
errors \citep[e.g.][]{2017MNRAS.469..721M,2021ARA&A..59...59B}. This means, based on pure proper motion 
data, dynamical modelings, which can only be achieved for nearby Sag-like systems at $\sim$50 to 60~kpc 
for the furthest now, is very promising to be applied to enough number of fainter tracers in dwarf systems 
at or beyond $\sim$100~kpc in the future. 

We now investigate the expected performance of using future CSST proper motions. The expected upper limit 
in CSST proper motion error ($\sim$0.2~mas/yr) is about a factor of two larger than that of {\it Gaia} DR3 at 
$18<g<19$. For very nearby systems such as Au16-9 at a distance a bit more than 20~kpc, the corresponding 
tangential velocity error is $\sim$10~km/s , which quickly increases to $\sim$40~km/s beyond 40~kpc. Thus it 
seems dynamical modeling with pure proper motion data is difficult with 0.2~mas/yr of error for systems more 
distant than $\sim$30$-$40~kpc. In Figure~\ref{fig:prof_infree_ml1_err_bkgd_csst}, we show the constraint 
based on the typical CSST proper motion error for Au16-9 as an example. The uncertainties are significantly 
larger, but the usage of pure proper motion data still leads to a reasonable fit.

The typical CSST proper motion error is based on about six astrometric measurements evenly distributed 
in 10 years of baseline. Our results thus indicate that this particular observational strategy would provide 
useful proper motions for dynamical modeling of very nearby systems at distances $<\sim$20~kpc. However, due 
to the fact that the CSST proper motion error estimates are based on the star number density in the Galactic 
bulge region, the actual errors based on the same operational mode can be smaller, so with pure CSST proper 
motion data, it could be possible to push slightly further. Notably, as will be shown in Nie et al., the 
associated proper motion errors increase slowly towards fainter magnitudes, reaching $\sim$0.4~mas/yr at 
CSST $g$-band apparent magnitude of 24-th, whereas {\it Gaia} proper motion errors increase very quickly 
beyond $G\sim20$. This is because the aperture and exposure time of CSST is designed to have higher 
signal-to-noise ratios at fainter magnitudes than {\it Gaia}, with the $g$-band limiting magnitude of 
about 26 to 27 \citep{zhan2011}. Thus future CSST proper motion measurements are promising to compensate 
{\it Gaia} measurements at fainter magnitudes.

\section{Discussions}
\label{sec:disc}
\subsection{Connection to the core-cusp problem}
\label{sec:corecusp}

\begin{figure} 
\includegraphics[width=0.49\textwidth]{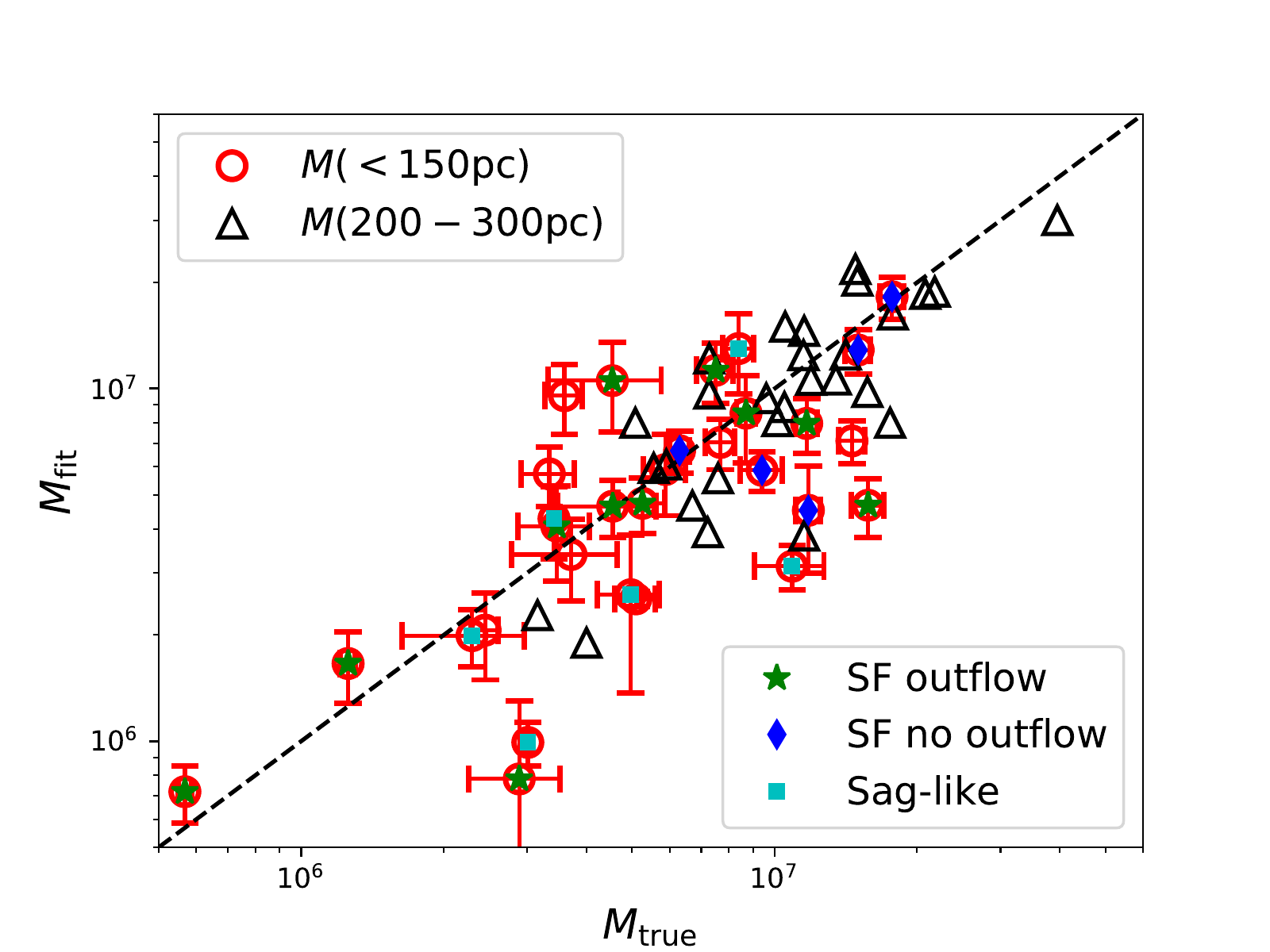}%
\caption{Red circles are besting-fitting versus true masses within 150~pc. The true masses are obtained 
by extrapolating the true density profiles to the very center based on the double power law functional 
form. he best-fitting mass within 150~pc is based on tracer dynamics beyond 185~pc, which is the softening 
scale of the simulation at $z=0$. Black empty triangles are best-fitting versus true masses between 200 and 
300~pc, which are exactly the same as those red circles in the left plot of Figure~\ref{fig:fit_vs_sf}. Red 
circles with a green star on top are star-forming dwarfs systems with prominent galactic winds. Red circles 
with a blue diamond on top are dwarfs with not zero but low star formation rates, and we do not see prominent
galactic winds for them. Red circles with a cyan square on top are nearby Sagittarius dSph-like systems. $y$ 
and $x$-axes errors reflect the 1-$\sigma$ uncertainties of the best fits and of the extrapolated 
inner profiles. Errors for the black empty triangles are comparable to or smaller than the symbol size, and 
thus are not shown. 
}
\label{fig:masstrue150_vs_massfit150}
\end{figure}

\begin{figure*} 
\includegraphics[width=0.49\textwidth]{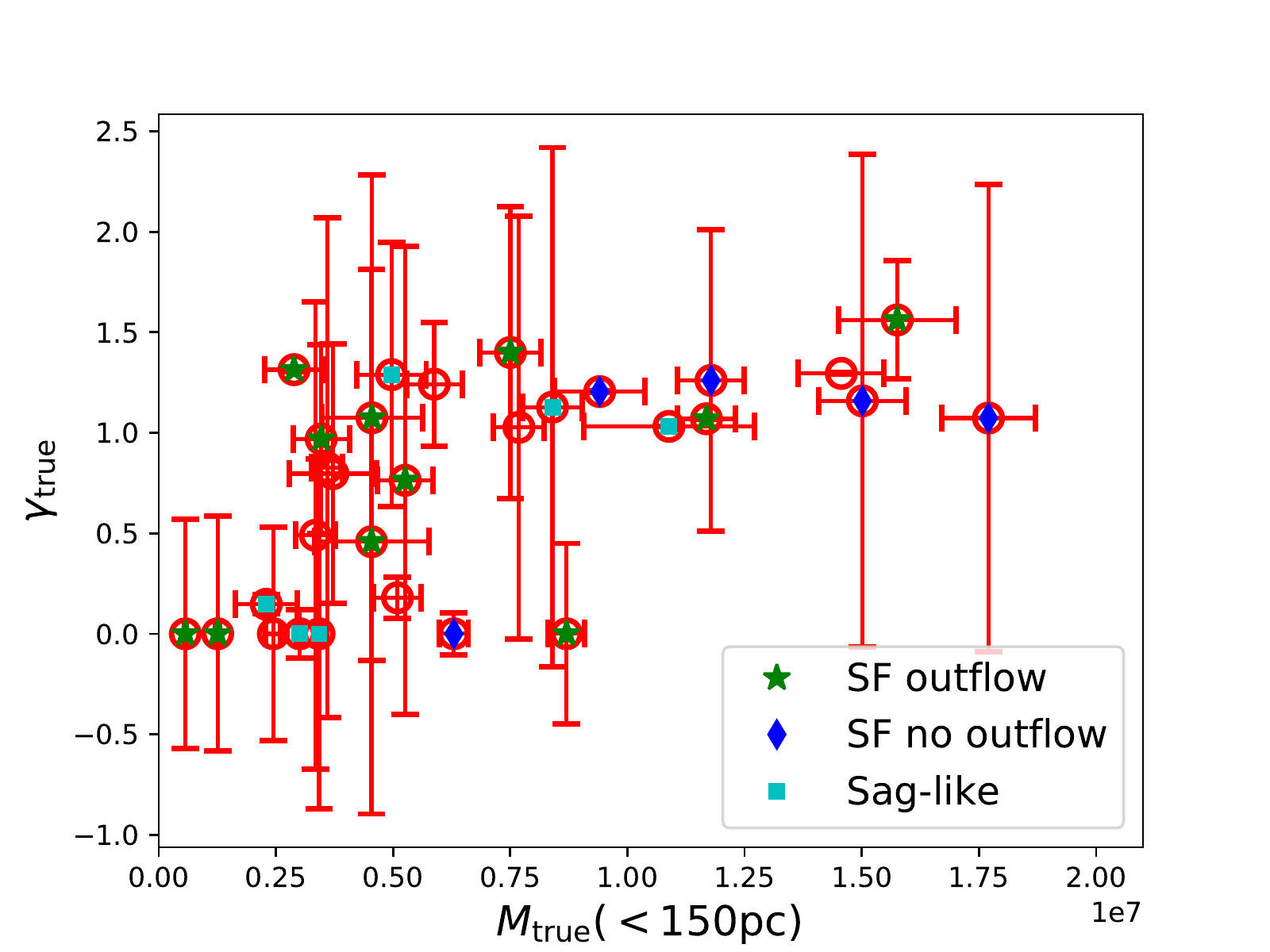}%
\includegraphics[width=0.49\textwidth]{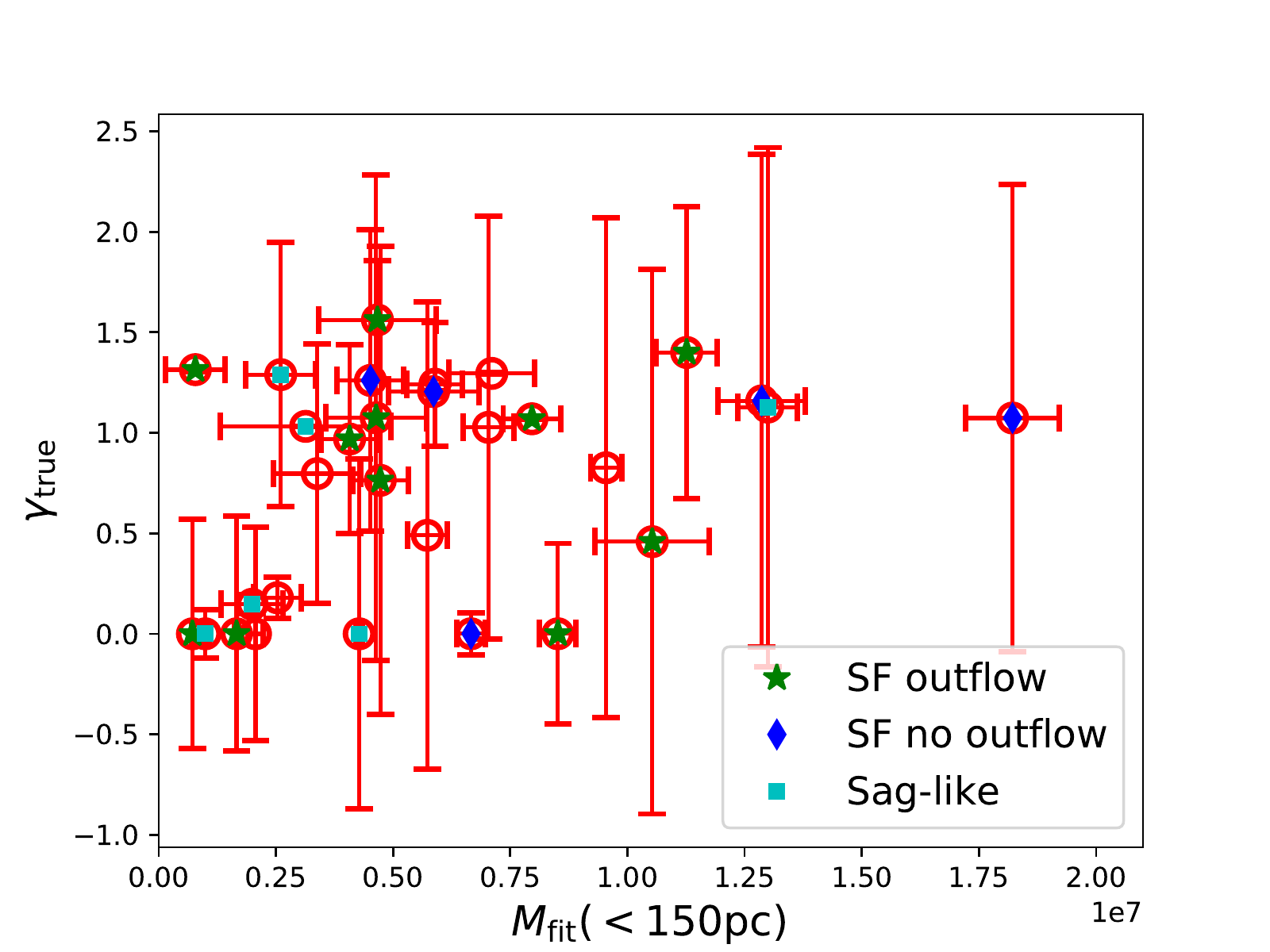}%
\caption{The true inner slopes versus the true (left) and best-fitting (right) masses enclosed within 
150~pc. The true inner slopes and true masses within 150~pc are computed by extrapolating the actual 
density profiles down to the very center, using a double power law function. The best-fitting mass 
within 150~pc is based on tracer dynamics beyond 185~pc. Meanings of different symbols are the same 
as Figure~\ref{fig:masstrue150_vs_massfit150}. 
}
\label{fig:slope_vs_mass}
\end{figure*}

The softening scale of the level 3 \textsc{auriga} set of simulations is 185~pc at $z=0$. For radial range below
this scale, the density profiles and dynamics of tracer particles are not realistic, so it is difficult for us to 
directly reach the scale that is the most relevant to the core-cusp problem \citep[150~pc; e.g.][]{2019MNRAS.484.1401R}.
However, based on our results in the previous section, the best-fitting inner density profiles show different 
amount of biases for individual systems, which also show a dependence on the star formation activities. 
It seems similar biases would exist on scales below the resolution limit as well, and thus we should be cautious 
when interpreting the results for individual systems. 

In Section~\ref{sec:profile}, the best-fitting inner profiles of a few quiescent Sag-like systems are shown to 
be below the true profiles when the inner slopes are allowed to be free. An interesting case is Au16-9. For this 
system, both the NFW and the double power law model profile lead to reasonable fits to the velocity map, but the 
qualitative goodness of fit is significantly better for the double power law model. Despite of the better fit 
to the tracer kinematics, the actual inner density profile is significantly under-estimated. Such an inconsistency 
between the kinematical prediction and the actual density distribution indicates strong deviations from steady states. 
Violation of the steady-state assumption could result in more core-ish inner profiles when the truth is more cuspy
for individual systems. Au16-9 is close to the galactic center and is undergoing some prominent contractions due 
to strong tidal effects, and we should avoid using systems which are strongly out of equilibrium, otherwise we may 
end up drawing wrong conclusions regarding the core-cusp issue, especially when looking at individual or small 
sample of systems. 

In real observation, instead of directly constraining the inner slopes, the best-fitting total mass 
within 150~pc, $M(<150\mathrm{pc})$, is often adopted to infer the steepness of the inner slope
\citep[e.g.][]{2019MNRAS.484.1401R}. 
We now compare the best-fitting and true mass within 150~pc. To achieve the comparison, we extrapolate the actual 
density profiles in the simulation down to scales below the resolution limit, based on the double power law model
profile. The extrapolation is used to calculate $M_\mathrm{true}(<150\mathrm{pc})$, which is regarded as the truth.
On the other hand, the best-fitting mass within 150~pc, $M_\mathrm{fit}(<150\mathrm{pc})$, is measured through tracer 
dynamics beyond 185~pc, extrapolated to the central region. Here we have excluded tracers within 185~pc to avoid 
wrong dynamics. Note in previous sections, our discussions are focused on the radial range not affected by star 
particles within 185~pc, and thus including or excluding tracers within 185~pc barely affects our results.  

The results are shown as red circles in Figure~\ref{fig:masstrue150_vs_massfit150}. The associated uncertainties
in the true masses are computed according to the boundaries which enclose 68\% of the MCMC chains\footnote{When fitting 
the double power law function to the true density profiles to achieve the extrapolation, we use \textsc{emcee}. }. 
Because there are no reliable tracers within 185~pc, now the \textsc{jam} constrained $M_\mathrm{fit}(<150\mathrm{pc})$ 
through stellar dynamics has significantly larger errorbars than those of $M_\mathrm{fit}(200-300\mathrm{pc})$ 
and $M_\mathrm{fit}(<r_\mathrm{half})$ in previous plots. For convenience, we also overplot the black empty triangles 
which are the measurements for $M(200-300\mathrm{pc})$. 

The scatter in $M(<150\mathrm{pc})$ is 0.255~dex, which is larger than the scatter of $M(200-300\mathrm{pc})$ 
(0.167~dex), but the overall fitting is reasonable. Star-forming and quiescent systems are well separated, 
with quiescent systems having smaller $M(<150\mathrm{pc})$. We believe the amount of scatter for red circles 
in Figure~\ref{fig:masstrue150_vs_massfit150} is an upper limit, because the scatter can be further 
decreased if tracers within 150~pc are available, and indeed in real observations stars within 150~pc 
can be observed and used for dynamical constraints.

We now check whether $M(<150\mathrm{pc})$ can be used as a good proxy to the inner slopes. Here we calculate 
the true inner slopes, $\gamma_\mathrm{true}$, based on the best-fitting parameter of the double power law 
model profile to the actual total density profiles beyond 185~pc in the simulation, so $\gamma_\mathrm{true}$
depends on extrapolations to the very center. Figure~\ref{fig:slope_vs_mass} shows $\gamma_\mathrm{true}$ 
versus both $M_\mathrm{fit}(<150\mathrm{pc})$ and $M_\mathrm{true}(<150\mathrm{pc})$. The scatters and errorbars 
are large, but there are correlations between $\gamma_\mathrm{true}$ and $M_\mathrm{true}(<150\mathrm{pc})$ or
$M_\mathrm{fit}(<150\mathrm{pc})$. In the left plot, $M_\mathrm{true}(<150\mathrm{pc})$ increases for steeper
$\gamma_\mathrm{true}$, indicating $M(<150\mathrm{pc})$ has the potential of being used as the proxy to infer 
the inner slopes. In the right plot, the correlation between $\gamma_\mathrm{true}$ and $M_\mathrm{fit}(<150\mathrm{pc})$ 
seems to be weakened due to the scatter in the recovery of $M(<150\mathrm{pc})$. However, we emphasize that since
we do not have tracers within 150~pc, the scatter in the right plot of Figure~\ref{fig:slope_vs_mass} and in 
Figure~~\ref{fig:masstrue150_vs_massfit150} is likely an upper limit. If tracers within 150~pc 
are available, the scatter is likely to be decreased. 

\subsection{The detectability of contractions and outflows}
\label{sec:detect}

\begin{figure*} 
\begin{center}
\includegraphics[width=0.9\textwidth]{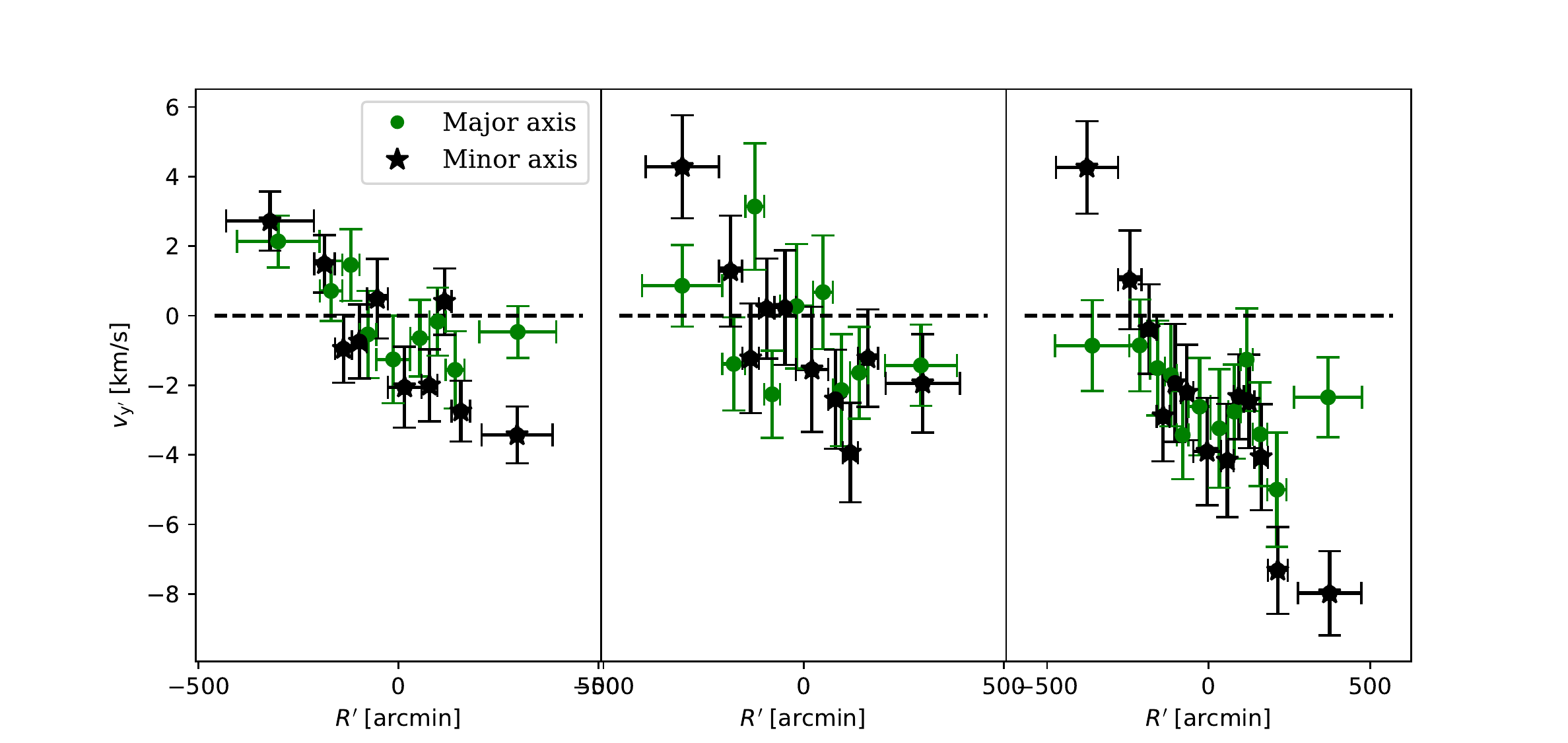}%
\end{center}
\caption{A demonstration of the detectability of contraction in nearby Sagittarius dSph-like systems. 
The figure is based on Au16-9, which is similar to the left plot of Figure~\ref{fig:xykin_nfw} but 
shows the $v_y'$ velocity component. Each bin contains 300 star particles. The left panel is based on bound 
member stars without including observational errors. Observational errors are incorporated in the middle 
panel. In the right panel, observational errors are included and tracers are selected according to their 
kinematics, which include some unbound particles in outskirts. We only show the $y'$-component of the 
tangential velocities along the major and minor axes of the dwarf. The black horizontal dashed lines 
of $y=0$ are to guide the eyes. }
\label{fig:xykin_detect} 
\end{figure*}

In Section~\ref{sec:results}, we discussed the contractions of member stars due to strong tidal effects for 
nearby systems, and star-forming systems are also mentioned to have outflows. Such systems are strongly out 
of equilibrium, and thus caution has to be taken if using them for dynamical modeling. Hence observational
identifications of such contractions and outflows are important. In this subsection, we briefly discuss the
detectability of such motions. 

To detect isotropic contractions or expansions, the usage of line-of-sight velocities of member stars or IFU 
observation might not be very useful, because the effect would be averaged and canceled out along the 
line-of-sight direction. In addition, if the contraction mainly happens along the image longer axis but not 
along the line-of-sight direction as in our case, line-of-sight velocities will not be helpful either. Proper 
motions thus become more important. 

In the left plot of Figure~\ref{fig:xykin_detect}, we show the contraction of member stars of Au16-9, without 
including any observational errors. It is clearly shown that $v_y'$ is negative/positive along the positive/negative 
$y'$-axis (minor axis), which demonstrates the contraction\footnote{Since along the major and minor axes, only 
stars within sectors of $\pm45$ degrees to the corresponding axes are used to make the plot, $v_y'$ binned along the 
$y'$ direction is dominated by radial motions. It is very prominent that $v_y'$ is negative along the positive minor 
axis, and positive along the negative minor axis (or $y'$-axis), which reflect inwards motions or contractions.}. 
Note the minor axis of Au16-9 is in fact the longer image axis, because our image major axis is defined according to 
the spin direction (see Section~\ref{sec:methods} for details), but the difference between the image major and minor 
axes of Au16-9 is very small anyway. Besides, $v_y'$ along the image major axis tends to show some rotations, but the 
rotation is absent in Figure~\ref{fig:xykin_nfw}, indicating Au16-9 could be triaxial. 

In the middle plot of Figure~\ref{fig:xykin_detect}, we include {\it Gaia} DR3 proper motion errors (see 
Section~\ref{sec:err_bkgd} for details), the signal is noisier but still clearly detectable. In the right plot, 
not only the proper motion errors are included, but also the stellar tracers are selected according to their 
difference in kinematics with respect to the dwarf center. The inclusion of some unbound particles in outskirts 
has caused some offsets towards negative $y$. This is because the global motion of the dwarf is defined through 
bound star particles and subtracted off from all selected tracer particles. However, those unbound particles are 
not expected to have zero mean velocity after the subtraction. Despite of the offset, the contraction is still 
detectable. 

Our results indicate, promisingly, that current {\it Gaia} DR3 proper motion errors can enable us to detect 
such contraction motions for member stars of nearby Sag-like systems. The detection is also benefited from the 
usage of a large sample of 6,000 member stars. 

\begin{figure} 
\includegraphics[width=0.49\textwidth]{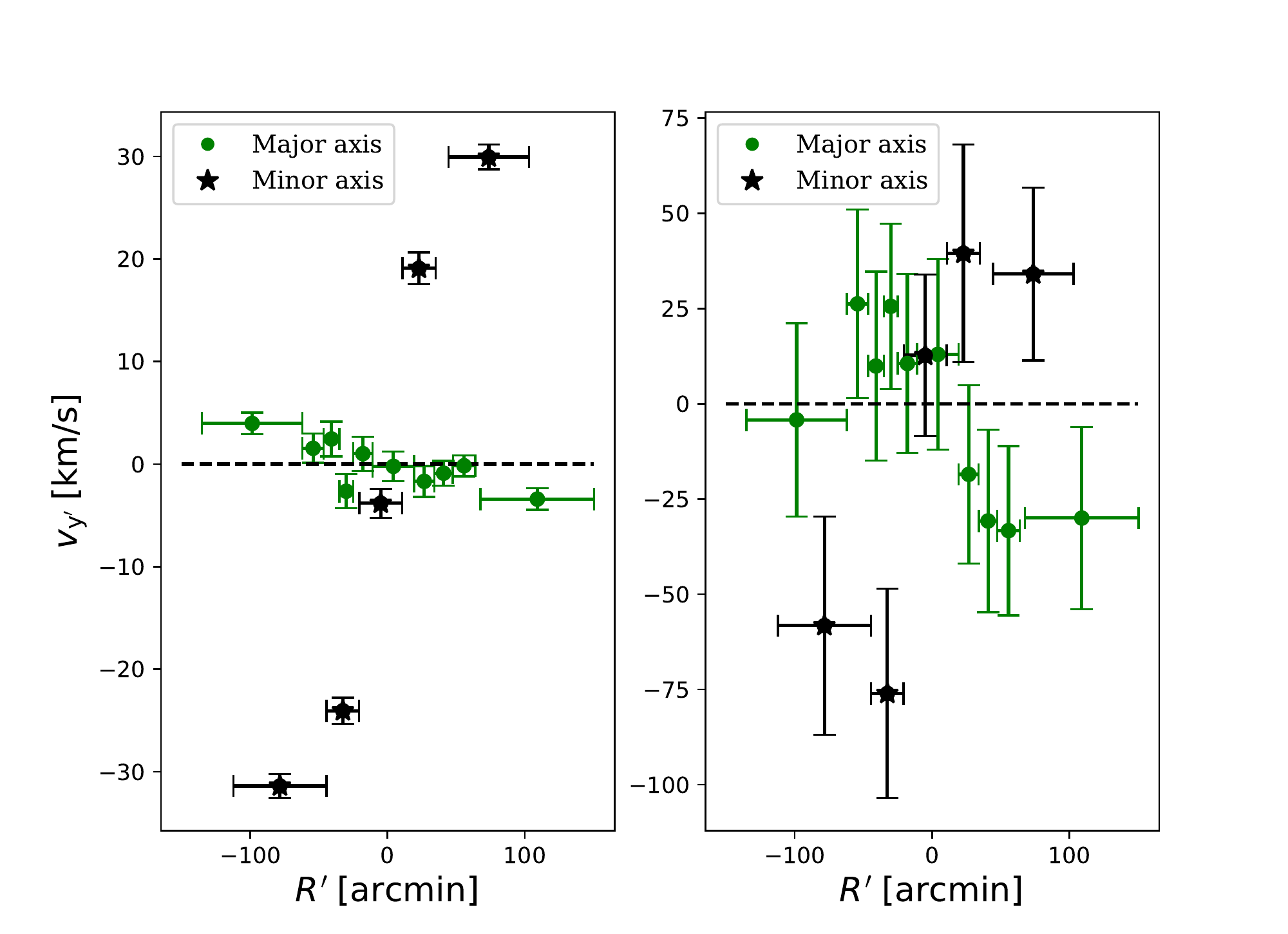}%
\caption{A demonstration of the detectability of outflows in a dwarf at a distance 
of $\sim$130~kpc. The figure is based on $\sim4,000$ wind particles in Au23-2. Each bin contains 
300 particles. Observational errors are included in the right plot, but not in the left plot. We 
only show the $y'$-component tangential velocities along major and minor axis of the dwarf, and 
the trend is the same for the $x'$-component tangential velocities. The black dots show prominent 
expansions, while the green dots show some rotations. The black horizontal dashed line at $y=0$ 
is to guide the eye.}
\label{fig:xykin_detect2}
\end{figure}

In contrast to the detection of global contractions in nearby systems, the detection of gas outflows can be more 
challenging. The 11 dwarfs with gas outflows are at fairly large distances. The most nearby dwarf 
with such outflow is at 130~kpc. Figure~\ref{fig:xykin_detect2} shows the major and minor axis proper motions of 
wind particles around this dwarf. In the left plot, no observational errors are incorporated, and we can clearly 
see the expansions, i.e., the $y'$ tangential velocities are positive at the positive $y'$-axis and negative at 
the negative $y'$-axis (black dots with errorbars). We now simply incorporate the {\it Gaia} DR3 proper motion 
errors to these wind particles in the right plot, treating them as individual stars. The trend becomes significantly 
noisy, but the signal is still marginally detectable for this system at 130~kpc. In real observation, the detection 
of such gas outflows perhaps has to depend on observations of the hot gas distribution around these dwarfs at 
different epochs.

\section{Conclusions}
\label{sec:concl}

In this study, we construct mock samples of member stars for 28 simulated dwarf galaxies from the cosmological 
level 3 set of \textsc{auriga} simulations. We applied the discrete axis-symmetric Jeans Anisotropic Multi-Gaussian 
Expansion (\textsc{jam}) modeling method to the 28 dwarf systems, to recover the underlying mass distribution through 
the stellar dynamics under the steady-state assumption. Among our sample of dwarf galaxies, there are 6 nearby
Sagittarius dSph-like (Sag-like) systems and other 11 star-forming systems with strong galactic winds or gas 
outflows.

We first use 6,000 bound member star particles without errors 
as stellar tracers for each system, which is to ensure good statistics. After assigning 
star particles apparent magnitudes using the multi-population synthesis \textsc{trilegal} code, we incorporate 
typical line-of-sight velocity errors (1$-$10~km/s), {\it Gaia} DR3 parallax errors and {\it Gaia} DR3 or China 
Space Station Telescope (CSST) proper motion errors. We also try a smaller sample of 2,000 star particles as 
tracers and try to select stellar tracers according to their differences in kinematics to the dwarf centers. 

Compared with the true matter distribution in the simulation, we find the recovered density profiles of the 
stellar and dark matter components individually are poor due to the degeneracy between stellar mass and 
dark matter distribution, and the constraints on the stellar-mass-to-light ratios are very weak. Fortunately, 
the density profiles of the total mass distribution can be constrained more reasonably. 

The total mass within the half-mass radius of tracers can be constrained the best, with a scatter of 
$\sim$0.067~dex. The best-fitting masses between 200 and 300~pc, $M_\mathrm{fit}(200-300\mathrm{pc})$, 
are recovered ensemble unbiasedly, with a scatter of 0.167 dex. However, the amount of bias shows 
a dependence on the current specific star formation rate (sSFR). This is a result of the deviation 
from steady states. Most quiescent Sag-like systems undergoing strong tidal effects have their
$M_\mathrm{fit}(200-300\mathrm{pc})$ under-estimated than the truth, whereas 
most star-forming systems with outflows have $M_\mathrm{fit}(200-300\mathrm{pc})$ over-estimated. 

The dependence of the best fits on sSFR reflects how tidal effects and galactic 
winds (gas outflows) perturb the system, driving it away from the steady state. We discover that most of 
the member stars in Sag-like systems have contractions along the image longer axis, which are the outcomes 
of strong tidal effects \citep[e.g.][]{2022MNRAS.510.2724O} and cannot be properly modeled by equilibrium 
models. On the other hand, the strong galactic winds or gas outflows in star-forming systems have driven 
the system out of equilibrium as well, despite the fact that wind particles themselves are not directly 
used as tracers. Tidal effects and galactic winds both cause the deviation from steady states, but in 
opposite ways, resulting in under-estimated and over-estimated inner density profiles, respectively. 
When talking about individual systems, one might end up with cored constraints on the inner profiles, 
whereas the truth is, on the contrary, more cuspy. On the other hand, cored systems might also be constrained 
as cuspy in the best fits for individual systems. Thus one should avoid drawing strong conclusions based 
on individual or limited number of systems. Instead, the averaged result based on a relatively large 
sample of dwarf galaxies is ensemble unbiased and more robust.

Interestingly, systems which still have some amount of star formation, but without prominent outflows,
mostly show smaller amount of biases in their best fits. Besides, massive quiescent systems at large  
distances and are not yet undergoing strong tidal effects, i.e., not Sag-like, also show smaller amount 
of biases on average. When using dynamical models based on the steady state assumption, we should be 
cautious when using systems undergoing prominent contractions or expansions, which are strongly out of
equilibrium. Promisingly, with {\it Gaia} DR3 proper motion errors, the contraction in nearby Sag-like 
systems is shown to be detectable. 

Using a smaller sample of 2,000 star particles with only line-of-sight velocities (typical errors from 1 to 
10~km/s) as tracers, $M(200-300\mathrm{pc})$ can still be constrained ensemble unbiasedly, but the scatter 
is increased by $\sim$50\%. 

After incorporating realistic observational errors (proper motion errors correspond to {\it Gaia} DR3) 
and selecting tracer stars based on the differences in their kinematics with respect to the dwarf center, 
we find their best constrained density profiles agree with the error-free constraints within 1-$\sigma$. 
For such nearby dwarf systems, dynamical constraints based on pure proper motions with {\it Gaia} DR3 
errors perform equally well or even better than the case of using pure line-of-sight velocities, which 
is very encouraging. On the other hand, the expected upper limit in proper motion errors for the future 
CSST survey is about twice larger than that of {\it Gaia} DR3, at CSST $g$-band apparent magnitudes of 
$18<g<19$, which can already lead to reasonable constraints for very nearby systems at distances 
$<\sim$20~kpc.

By extrapolating the true density profiles in the simulation down to the radial range below the resolution 
limit, and by extrapolating the best-fitting density profiles based on tracer dynamics at $r>185$~pc down 
to the center, we find the mass within 150~pc can be recovered ensemble unbiasedly as well, but with a 
larger scatter of 0.255~dex. We also see correlations between the best-fitting or true mass within 150~pc 
and the true inner slopes, but with large scatters. The scatter can be improved if tracers within 150~pc
are available.

\acknowledgments

We acknowledge the science research grants from the China Manned Space (CSST) Project with 
NO. CMS-CSST-2021-B03. This work is also supported in part by the CSST Project grant No. 
CMS-CSST-2021-A02, NSFC (12022307), the National Key Basic Research and Development 
Program of China (No.2018YFA0404504), 111 project (No. B20019) and Shanghai Natural Science 
Foundation (No. 19ZR1466800). We thank the sponsorship from Yangyang Development Fund. WW 
is grateful for useful discussions with Xiaoting Fu, Leo Girardi, Zhiyuan Li, Lu Li, Fabo Feng 
and Zhen Yuan. The computation of this work is carried out on the \textsc{Gravity} supercomputer 
at the Department of Astronomy, Shanghai Jiao Tong University, and is partly supported by 
the STFC DiRAC HPC Facility, at the Institute of Computational Cosmology (ICC), Durham 
University. LZ acknowledge funding from the National Key R\&D Program of China under grant 
No. 2018YFA0404501, and National Natural Science Foundation of China under grant No. Y945271001.
RG acknowledges financial support from the Spanish Ministry of Science and Innovation 
(MICINN) through the Spanish State Research Agency, under the Severo Ochoa Program 
2020-2023 (CEX2019-000920-S). FAG acknowledges support from ANID FONDECYT Regular 
1211370 and by the ANID BASAL project FB210003. FAG, and acknowledges funding from the 
Max Planck Society through a “Partner Group” grant.

%







\bibliography{master}{}
\bibliographystyle{aasjournal}

\clearpage



\end{document}